\documentstyle[12pt]{book} 
\begin{document}

\sloppy
\raggedbottom

\begin{titlepage}

\hfill\vspace{1in}\\
{\Huge\bf\hspace{-\parindent}The Limits of \vspace{3mm}\\
Mathematics--- \vspace{3mm} \\
\rm Tutorial Version \vspace{1in} \\
}
{\Large\it To Fran\c{c}oise \vspace{1in} \\
\bf G J Chaitin \vspace{1mm} \\
IBM, P O Box 704 \\
Yorktown Heights, NY 10598 \vspace{1mm} \\
{\it chaitin@watson.ibm.com} \vspace{1in} \\
September 12, 1995
}

\end{titlepage}

\begin{titlepage}

\section*{Abstract}
The latest in a series of reports presenting the information-theoretic
incompleteness theorems of algorithmic information theory via
algorithms written in specially designed versions of LISP.  Previously
in this LISP code only one-character identifiers were allowed, and
arithmetic had to be programmed out.  Now identifiers can be many
characters long, and arithmetic with arbitrarily large unsigned
decimal integers is built in.  This and many other changes in the
software have made this material much easier to understand and to use.

\end{titlepage}

\markboth
{The Limits of Mathematics}{Tutorial Version}

\chapter*{Preface}

In a remarkable development, I have constructed a new definition for a
self-delimiting universal Turing machine (UTM) that is easy to program
and runs very quickly.  This provides a new foundation for algorithmic
information theory (AIT), which is the theory of the size in bits of
programs for self-delimiting UTM's.  Previously, AIT had an abstract
mathematical quality.  Now it is possible to write down executable
programs that embody the constructions in the proofs of theorems.  So
AIT goes from dealing with remote idealized mythical objects to being
a theory about practical down-to-earth gadgets that one can actually
play with and use.

This new self-delimiting UTM is implemented via software written in a
new version of LISP that I invented especially for this purpose.  This
LISP was designed by writing a interpreter for it in Mathematica that
was then translated into C\@. I have tested this software by running it on
IBM RS/6000 workstations with the AIX version of UNIX.

Using this new software and the latest theoretical ideas, it
is now possible to give a self-contained ``hands on'' mini-course
presenting very concretely my latest proofs of my two fundamental
information-theoretic incompleteness theorems.  The first of these
theorems states that an $N$-bit formal axiomatic system cannot enable
one to exhibit any specific object with program-size complexity
greater than $N+c$.  The second of these theorems states that an
$N$-bit formal axiomatic system cannot enable one to determine more
than $N+c'$ scattered bits of the halting probability $\Omega$.

Most people believe that anything that is true is true for a reason.
These theorems show that some things are true for no reason at all,
i.e., accidentally, or at random.

As is shown in this course, the algorithms considered in the
proofs of these two theorems are now easy to program and run, and by
looking at the size in bits of these programs one can actually, for
the first time, determine exact values for the constants $c$ and $c'$.

I used this approach and software in an intensive short course on the
limits of mathematics that I gave at the University of Maine in Orono
in the summer of 1994.  I also lectured on this material during a
stay at the Santa Fe Institute in the spring of 1995, and at a meeting
at the Black Sea University in Romania in the summer of 1995.  A
summary of the approach that I used on these three occasions will
appear under the title ``A new version of algorithmic information
theory'' in a forthcoming issue of the new
magazine {\it Complexity,} which has just
been launched by the Santa Fe Institute and John Wiley and Sons.

After presenting this material at these three different places, it
became obvious to me that it is extremely difficult to understand it
in its original form.  So next time, at the Rovaniemi Institute of
Technology in the summer of 1996, I am going to use the new, more
understandable software in this report; everything has been redone in
an attempt to make it as easy to understand as possible.

For their stimulating invitations, I thank Prof.\ George Markowsky of
the University of Maine, Prof.\ Cristian Calude of the
University of Auckland, Prof.\ John Casti of the Santa Fe Institute,
and Prof.\ Veikko Ker\"anen of the Rovaniemi Institute of Technology.
And I am grateful to IBM for supporting my research for almost thirty
years, and to my current management chain at the IBM Research
Division, Dan Prener, Christos Georgiou, Eric Kronstadt, Jeff Jaffe,
and Jim McGroddy.

All enquires, comments and suggestions regarding this software should
be sent via e-mail to {\tt chaitin@watson.ibm.com}.

\tableofcontents

\newcommand
{\chap}[1]{\chapter*{#1}\markboth{The Limits of Mathematics}{#1}
\addcontentsline{toc}{chapter}{#1}}
\newcommand
{\Size}{\tiny}

\part{Explanation}

\chap{The New Idea}

Here is a quick summary of this new LISP, in which atoms can now either be
words or unsigned decimal integers.  First of all, comments are
written like this: {\tt [comment]}.  Each LISP primitive function has
a fixed number of arguments.  {\tt '} is {\tt QUOTE}, {\tt =} is {\tt
EQ}, and {\tt atom, car, cdr, cadr, caddr, cons} are provided with
their usual meaning.  We also have {\tt lambda, define, let, if} and
{\tt display} and {\tt eval}.  The notation {\tt "} indicates that an
S-expression with explicit parentheses follows, not what is usually
the case in this LISP, an M-expression, in which the parentheses for
each primitive function are implicit.  {\tt nil} denotes the empty
list {\tt ()}, and the logical truth values are {\tt true} and {\tt
false}.  For dealing with unsigned decimal integers we have \verb|+,
-, *, ^, <, >, <=, >=, base10-to-2, base2-to-10|.

So far this is fairly standard.  The new idea is this.  We define our
standard self-delimiting universal Turing machine as follows.  Its
program is in binary, and appears on a tape in the following form.
First comes a LISP expression, written in ASCII with 8 bits per
character, and terminated by an end-of-line character \verb|'\n'|.  The
TM reads in this LISP expression, and then evaluates it.  As it does
this, two new primitive functions {\tt read-bit} and {\tt read-exp}
with no arguments may be used to read more from the TM tape.  Both of
these functions explode if the tape is exhausted, killing the
computation.  {\tt read-bit} reads a single bit from the tape, and
{\tt read-exp} reads in an entire LISP expression, in 8-bit character
chunks.

This is the only way that information on the TM tape may be accessed,
which forces it to be used in a self-delimiting fashion.  This is
because no algorithm can search for the end of the tape and then use
the length of the tape as data in the computation.  If an algorithm
attempts to read a bit that is not on the tape, the algorithm aborts.

How is information placed on the TM tape in the first place?  Well, in
the starting environment, the tape is empty and any attempt to read it
will give an error message.  To place information on the tape, one
must use the primitive function {\tt try} which tries to see if an
expression can be evaluated.

Consider the three arguments $\alpha$, $\beta$ and $\gamma$ of {\tt
try}.  The meaning of the first argument is as follows.  If $\alpha$
is {\tt no-time-limit}, then there is no depth limit.  Otherwise
$\alpha$ must be an unsigned decimal integer, and gives the depth
limit (limit on the nesting depth of function calls and
re-evaluations).  The second argument $\beta$ of {\tt try} is the
expression to be evaluated as long as the depth limit $\alpha$ is not
exceeded.  And the third argument $\gamma$ of {\tt try} is a list of
bits to be used as the TM tape.

The value $\nu$ returned by the primitive function {\tt try} is a
triple.  The first element of $\nu$ is {\tt success} if the evaluation
of $\beta$ was completed successfully, and the first element of $\nu$
is {\tt failure} if this was not the case.  The second element of
$\nu$ is {\tt out-of-data} if the evaluation of $\beta$ aborted
because an attempt was made to read a non-existent bit from the TM
tape.  The second element of $\nu$ is {\tt out-of-time} if evaluation
of $\beta$ aborted because the depth limit $\alpha$ was exceeded.
These are the only possible error flags, because this LISP is designed
with maximally permissive semantics.  If the computation $\beta$
terminated normally instead of aborting, the second element of $\nu$
will be the result produced by the computation $\beta$, i.e., its
value.  That's the second element of the list $\nu$ produced by the
{\tt try} primitive function.

The third element of the value $\nu$ is a list of all the arguments to
the primitive function {\tt display} that were encountered during the
evaluation of $\beta$.  More precisely, if {\tt display} was called
$N$ times during the evaluation of $\beta$, then $\nu$ will be a list
of $N$ elements.  The $N$ arguments of {\tt display} appear in $\nu$
in chronological order.  Thus {\tt try} can not only be used to
determine if a computation $\beta$ reads too much tape or goes on too
long (i.e., to greater depth than $\alpha$), but {\tt try} can also be
used to capture all the output that $\beta$ displayed as it went
along, whether the computation $\beta$ aborted or not.

In summary, all that one has to do to simulate a self-delimiting
universal Turing machine $U(p)$ running on the binary program $p$ is
to write
\begin{verbatim}
          try no-time-limit 'eval read-exp p
\end{verbatim}
This is an M-expression with parentheses omitted from primitive
functions.  (Recall that all primitive functions have a fixed number
of arguments.)  With the parentheses supplied, it becomes the
S-expression
\begin{verbatim}
          (try no-time-limit ('(eval(read-exp))) p)
\end{verbatim}
This says that one is to read a complete LISP S-expression from the TM
tape $p$ and then evaluate it without any time limit and using
whatever is left on the tape $p$.

Some more primitive functions have also been added.  The 2-argument
function {\tt append} denotes list concatenation, and the 1-argument
function {\tt bits} converts an S-expression into the list of the bits
in its ASCII character string representation.  These are used for
constructing the bit strings that are then put on the TM tape using
{\tt try}'s third argument $\gamma$.  We also provide the 1-argument
functions {\tt size} and {\tt length} that respectively give the
number of characters in an S-expression, and the number of elements in
a list.  Note that the functions {\tt append}, {\tt size} and {\tt
length} could be programmed rather than included as built-in primitive
functions, but it is extremely convenient and much much faster to
provide them built in.

Finally a new 1-argument identity function {\tt debug} with the
side-effect of outputting its argument is provided for debugging.
Output produced by {\tt debug} is invisible to the ``official'' {\tt
display} and {\tt try} output mechanism.  {\tt debug} is needed
because {\tt try} $\alpha$ $\beta$ $\gamma$ suppresses all output
$\theta$ produced within its depth-controlled evaluation of $\beta$.
Instead {\tt try} collects all output $\theta$ from within $\beta$ for
inclusion in the final value $\nu$ that {\tt try} returns, namely $\nu
= $ (success/failure, value of $\beta$, $\theta$).

\chap{Course Outline}

The course begins by explaining with examples my new LISP.  See {\tt
examples.r}.

Then the theory of LISP program-size complexity is developed a little
bit.  LISP program-size complexity is extremely simple and concrete.
In particular, it is easy to show that it is impossible to prove that
a self-contained LISP expression is elegant, i.e., that no smaller
expression has the same value.  To prove that an $N$-character LISP
expression is elegant requires a formal axiomatic system that itself
has at least LISP complexity $N-418$.  See {\tt godel.r}.

Next we define our standard self-delimiting universal Turing machine
$U(p)$ using
\begin{verbatim}
          cadr try no-time-limit 'eval read-exp p
\end{verbatim}
as explained in the previous chapter.

Next we show that
\[
   H(x,y) \le H(x) + H(y) + c
\]
with $c = 432.$
Here $H(\cdots)$ denotes the size in bits of the smallest program that
makes our standard universal Turing machine compute $\cdots$.  Thus
this inequality states that the information needed to compute the pair
$(x,y)$ is bounded by a constant $c$ plus the sum of the information
needed to compute $x$ and the information needed to compute $y$.
Consider
\begin{verbatim}
          cons eval read-exp
          cons eval read-exp
               nil
\end{verbatim}
This is an M-expression with parentheses omitted from primitive
functions.  With all the parentheses supplied, it becomes the S-expression
\begin{verbatim}
          (cons (eval (read-exp))
          (cons (eval (read-exp))
                nil))
\end{verbatim}
$c = 432$ is just 8 bits plus 8 times the size in characters of this LISP
S-expression.  See {\tt univ.r}.

Consider a binary string $x$ whose size is $|x|$ bits.  In {\tt
univ.r} we also show that
\[
   H(x) \le 2|x| + c
\]
and
\[
   H(x) \le |x| + H(|x|) + c'
\]
with $c = 1106$ and $c' = 1152$.  As before, the programs for doing
this are exhibited and run.

Next we turn to the self-delimiting program-size complexity $H(X)$ for
infinite r.e.\ sets $X$.  This is defined to be the size in bits of
the smallest LISP expression $\xi$ that executes forever without
halting and outputs the members of the r.e.\ set $X$ using the LISP
primitive {\tt display}, which is an identity function with the
side-effect of outputting the value of its argument.  Note that this
LISP expression $\xi$ is allowed to read additional bits or
expressions from the TM tape using the primitive functions {\tt
read-bit} and {\tt read-exp} if $\xi$ so desires.  But of course $\xi$
is charged for this; this adds to $\xi$'s program size.

It is in order to deal with such unending expressions $\xi$ that the
LISP primitive function for time-limited evaluation {\tt try} captures
all output from {\tt display} within its second argument $\beta$.

Now consider a formal axiomatic system $A$ of complexity $N$, i.e.,
with a set of theorems $T_A$ that considered as an r.e.\ set as above
has self-delimiting program-size complexity $H(T_A) = N$.  We show
that $A$ cannot enable us to exhibit a specific S-expression $s$ with
self-delimiting complexity $H(s)$ greater than $N+c$.  Here $c =
4696$.  See {\tt godel2.r}.

Next we show two different ways to calculate the halting probability
$\Omega$ of our standard self-delimiting universal Turing machine in
the limit from below.  See {\tt omega.r} and {\tt omega2.r}.  The
first LISP program for doing this, {\tt omega.l}, is quite
straight-forward.  The second program for calculating $\Omega$, {\tt
omega2.l}, uses a much more clever method than {\tt omega.l} does.
Using {\tt omega2.l} as a subroutine, we show that if $\Omega_N$
is the first $N$ bits of the fractional part of the base-two real
number $\Omega$, then
\[
   H(\Omega_N) > N - c
\]
with $c = 9488$.  Again this is done with a program that can actually
be run and whose size gives us a value for $c$.  See {\tt omega3.r}.

Consider again the formal axiomatic system $A$ with complexity $N$,
i.e., with self-delimiting program-size complexity $H(T_A) = N$.
Using the lower bound of $N-c$ on $H(\Omega_N)$ established in {\tt
omega3.r}, we show that $A$ cannot enable us to determine more than
the first $N+c'$ bits of $\Omega$.  Here $c' = 16400$.  In fact, we
show that $A$ cannot enable us to determine more than $N+c'$ bits of
$\Omega$ even if they are scattered and we leave gaps.  See {\tt
godel3.r}.

Last but not least, the philosophical implications of all this
should be discussed, especially the extent to which it tends to
justify experimental mathematics.  This would be along the lines of
the discussion in my talk transcript ``Randomness in arithmetic and
the decline and fall of reductionism in pure mathematics,'' in J.
Cornwell, {\it Nature's Imagination,} Oxford University Press, 1995,
pp.\ 27--44.

This concludes our ``hand-on'' mini-course on the
information-theoretic limits of mathematics.

\chap{Software User Guide}

All the software for this course is written in a new version of {\sl
LISP}.  The interpreter for this new {\sl LISP} was originally written
in {\sl Mathematica}.  See {\tt lisp.m}.  Then the interpreter was
re-written in {\sl C}\@.  See {\tt lisp.c}.

I used Version 2.1 of {\sl Mathematica} as described in the second
edition of Wolfram's book {\it Mathematica---A System for Doing
Mathematics by Computer,} running on {\sl IBM RISC} System/6000
workstations.

There are four different kinds of files:

\begin{enumerate}
\item {\tt *.m} files are {\sl Mathematica} code.
\item {\tt *.c} files are {\sl C} code.
\item {\tt *.l} files are {\sl LISP} code.
\item {\tt *.r} files are the output from {\sl LISP} runs.
\end{enumerate}

To run the {\tt lisp.m} interpreter, first enter {\sl Mathematica}
using the command {\tt math}.  Then tell {\sl Mathematica},
\begin{verbatim}
               << lisp.m
\end{verbatim}
To run a {\sl LISP} program {\tt xyz.l} and produce output file {\tt
xyz.r}, enter
\begin{verbatim}
               run @ "xyz"
\end{verbatim}
To run several programs, enter
\begin{verbatim}
               run /@ {"xxx","yyy","zzz"}
\end{verbatim}
To run the eight {\sl LISP} program in the course, enter
\begin{verbatim}
               runall
\end{verbatim}
Type {\tt Exit} to exit from {\sl Mathematica}.

Here is how to run the programs that compute the halting probability
$\Omega$ in the limit from below:
\begin{verbatim}
               math
               << lisp.m
               run /@ {"omega","omega2"}
               run @ "omega3"
               Exit
\end{verbatim}

To run the {\tt lisp.c} interpreter, first compile it with the command
\begin{verbatim}
               cc -O -olisp lisp.c
\end{verbatim}
This produces a 32 megabyte interpreter.  If that is too large, reduce
the \verb|#define SIZE 1000000| in {\tt lisp.c} and recompile it.  To
run a {\sl LISP} program {\tt xyz.l} and produce output file {\tt xyz.r},
enter
\begin{verbatim}
               lisp < xyz.l > xyz.r
\end{verbatim}

These two {\sl LISP} interpreters run at vastly different speeds, but
they should always produce identical results.  This can easily be
checked, for example as follows:
\begin{verbatim}
               diff xyz.r xyz.r2 > out
               vi out
\end{verbatim}

\chap{Bibliography}

\begin{itemize}
\item[{[1]}] S. Wolfram, {\it Mathematica---A System for Doing
Mathematics by Computer,} second edition, Addison-Wesley, 1991.
\item[{[2]}] B. W. Kernighan and D. M. Ritchie, {\it The C Programming
Language,} second edition, Prentice Hall, 1988.
\item[{[3]}] G. J. Chaitin, ``Randomness in arithmetic and the decline
and fall of reductionism in pure mathematics,'' in J. Cornwell, {\it
Nature's Imagination,} Oxford University Press, 1995, pp.\ 27--44.
\item[{[4]}] G. J. Chaitin, ``The Berry paradox,'' {\it Complexity\/}
1 (1995), pp.\ 26--30.
\item[{[5]}] G. J. Chaitin, ``A new version of algorithmic information
theory,'' {\it Complexity,} to appear.
\end{itemize}

\part{The Course}

\chap{examples.r}{\Size\begin{verbatim}
lisp.c

LISP Interpreter Run

[ test new lisp ]
' (ab c d)

expression  (' (ab c d))
value       (ab c d)

'(ab   cd  )

expression  (' (ab cd))
value       (ab cd)

car '(aa bb cc)

expression  (car (' (aa bb cc)))
value       aa

cdr '(aa bb cc)

expression  (cdr (' (aa bb cc)))
value       (bb cc)

cadr '(aa bb cc)

expression  (car (cdr (' (aa bb cc))))
value       bb

caddr '(aa bb cc)

expression  (car (cdr (cdr (' (aa bb cc)))))
value       cc

cons '(aa bb cc) '(dd ee ff)

expression  (cons (' (aa bb cc)) (' (dd ee ff)))
value       ((aa bb cc) dd ee ff)

car aa

expression  (car aa)
value       aa

cdr aa

expression  (cdr aa)
value       aa

cons aa bb

expression  (cons aa bb)
value       aa

("cons aa)

expression  (cons aa)
value       (aa)

("cons '(aa) '(bb) '(cc))

expression  (cons (' (aa)) (' (bb)) (' (cc)))
value       ((aa) bb)

let x a x

expression  ((' (lambda (x) x)) a)
value       a

x

expression  x
value       x

atom ' aa

expression  (atom (' aa))
value       true

atom '(aa)

expression  (atom (' (aa)))
value       false

if true x y

expression  (if true x y)
value       x

if false x y

expression  (if false x y)
value       y

if xxx x y

expression  (if xxx x y)
value       x

let (f x) if atom display x x (f car x)
 (f '(((a)b)c))

expression  ((' (lambda (f) (f (' (((a) b) c))))) (' (lambda (
            x) (if (atom (display x)) x (f (car x))))))
display     (((a) b) c)
display     ((a) b)
display     (a)
display     a
value       a

f

expression  f
value       f

let (cat x y) if atom x y cons car x (cat cdr x y)
    (cat '(a b c) '(d e f))

expression  ((' (lambda (cat) (cat (' (a b c)) (' (d e f)))))
            (' (lambda (x y) (if (atom x) y (cons (car x) (cat
             (cdr x) y))))))
value       (a b c d e f)

cat

expression  cat
value       cat

define (cat x y) if atom x y cons car x (cat cdr x y)

define      cat
value       (lambda (x y) (if (atom x) y (cons (car x) (cat (c
            dr x) y))))

cat

expression  cat
value       (lambda (x y) (if (atom x) y (cons (car x) (cat (c
            dr x) y))))

(cat '(a b c) '(d e f))

expression  (cat (' (a b c)) (' (d e f)))
value       (a b c d e f)

define x cadr '(a b c)

expression  (car (cdr (' (a b c))))
define      x
value       b

x

expression  x
value       b

define x caddr '(a b c)

expression  (car (cdr (cdr (' (a b c)))))
define      x
value       c

x

expression  x
value       c

length display
bits ' a

expression  (length (display (bits (' a))))
display     (0 1 1 0 0 0 0 1 0 0 0 0 1 0 1 0)
value       16

length display
bits ' abc

expression  (length (display (bits (' abc))))
display     (0 1 1 0 0 0 0 1 0 1 1 0 0 0 1 0 0 1 1 0 0 0 1 1 0
             0 0 0 1 0 1 0)
value       32

nil

expression  nil
value       ()

length display
bits nil

expression  (length (display (bits nil)))
display     (0 0 1 0 1 0 0 0 0 0 1 0 1 0 0 1 0 0 0 0 1 0 1 0)
value       24

length display
bits ' (a)

expression  (length (display (bits (' (a)))))
display     (0 0 1 0 1 0 0 0 0 1 1 0 0 0 0 1 0 0 1 0 1 0 0 1 0
             0 0 0 1 0 1 0)
value       32

size abc

expression  (size abc)
value       3

size ' ( a b c )

expression  (size (' (a b c)))
value       7

length ' ( a b c )

expression  (length (' (a b c)))
value       3

+ abc 15

expression  (+ abc 15)
value       15

+ '(abc) 15

expression  (+ (' (abc)) 15)
value       15

+ 10 15

expression  (+ 10 15)
value       25

- 10 15

expression  (- 10 15)
value       0

- 15 10

expression  (- 15 10)
value       5

* 10 15

expression  (* 10 15)
value       150

^ 10 15

expression  (^ 10 15)
value       1000000000000000

< 10 15

expression  (< 10 15)
value       true

< 10 10

expression  (< 10 10)
value       false

> 10 15

expression  (> 10 15)
value       false

> 10 10

expression  (> 10 10)
value       false

<= 10 15

expression  (<= 10 15)
value       true

<= 10 10

expression  (<= 10 10)
value       true

>= 10 15

expression  (>= 10 15)
value       false

>= 10 10

expression  (>= 10 10)
value       true

= 10 15

expression  (= 10 15)
value       false

= 10 10

expression  (= 10 10)
value       true

let (f x) if = 0 x 1 * display x (f - x 1)
    (f 5)

expression  ((' (lambda (f) (f 5))) (' (lambda (x) (if (= 0 x)
             1 (* (display x) (f (- x 1)))))))
display     5
display     4
display     3
display     2
display     1
value       120

let (f x) if = 0 x 1 * x (f - x 1)
    (f 100)

expression  ((' (lambda (f) (f 100))) (' (lambda (x) (if (= 0
            x) 1 (* x (f (- x 1)))))))
value       93326215443944152681699238856266700490715968264381
            62146859296389521759999322991560894146397615651828
            62536979208272237582511852109168640000000000000000
            00000000

try 0
'let (f x) if = 0 x 1 * display x (f - x 1)
     (f 5)
nil

expression  (try 0 (' ((' (lambda (f) (f 5))) (' (lambda (x) (
            if (= 0 x) 1 (* (display x) (f (- x 1)))))))) nil)
value       (failure out-of-time ())

try 1
'let (f x) if = 0 x 1 * display x (f - x 1)
     (f 5)
nil

expression  (try 1 (' ((' (lambda (f) (f 5))) (' (lambda (x) (
            if (= 0 x) 1 (* (display x) (f (- x 1)))))))) nil)
value       (failure out-of-time ())

try 2
'let (f x) if = 0 x 1 * display x (f - x 1)
     (f 5)
nil

expression  (try 2 (' ((' (lambda (f) (f 5))) (' (lambda (x) (
            if (= 0 x) 1 (* (display x) (f (- x 1)))))))) nil)
value       (failure out-of-time (5))

try 3
'let (f x) if = 0 x 1 * display x (f - x 1)
     (f 5)
nil

expression  (try 3 (' ((' (lambda (f) (f 5))) (' (lambda (x) (
            if (= 0 x) 1 (* (display x) (f (- x 1)))))))) nil)
value       (failure out-of-time (5 4))

try 4
'let (f x) if = 0 x 1 * display x (f - x 1)
     (f 5)
nil

expression  (try 4 (' ((' (lambda (f) (f 5))) (' (lambda (x) (
            if (= 0 x) 1 (* (display x) (f (- x 1)))))))) nil)
value       (failure out-of-time (5 4 3))

try 5
'let (f x) if = 0 x 1 * display x (f - x 1)
     (f 5)
nil

expression  (try 5 (' ((' (lambda (f) (f 5))) (' (lambda (x) (
            if (= 0 x) 1 (* (display x) (f (- x 1)))))))) nil)
value       (failure out-of-time (5 4 3 2))

try 6
'let (f x) if = 0 x 1 * display x (f - x 1)
     (f 5)
nil

expression  (try 6 (' ((' (lambda (f) (f 5))) (' (lambda (x) (
            if (= 0 x) 1 (* (display x) (f (- x 1)))))))) nil)
value       (failure out-of-time (5 4 3 2 1))

try 7
'let (f x) if = 0 x 1 * display x (f - x 1)
     (f 5)
nil

expression  (try 7 (' ((' (lambda (f) (f 5))) (' (lambda (x) (
            if (= 0 x) 1 (* (display x) (f (- x 1)))))))) nil)
value       (success 120 (5 4 3 2 1))

try no-time-limit
'let (f x) if = 0 x 1 * display x (f - x 1)
     (f 5)
nil

expression  (try no-time-limit (' ((' (lambda (f) (f 5))) (' (
            lambda (x) (if (= 0 x) 1 (* (display x) (f (- x 1)
            ))))))) nil)
value       (success 120 (5 4 3 2 1))

eval display '+ 5 15

expression  (eval (display (' (+ 5 15))))
display     (+ 5 15)
value       20

try 6
'let (f x) if = 0 x nil
           cons * 2 display read-bit (f - x 1)
     (f 5)
'(1 0 1 0 1)

expression  (try 6 (' ((' (lambda (f) (f 5))) (' (lambda (x) (
            if (= 0 x) nil (cons (* 2 (display (read-bit))) (f
             (- x 1)))))))) (' (1 0 1 0 1)))
value       (failure out-of-time (1 0 1 0 1))

try 7
'let (f x) if = 0 x nil
           cons * 2 display read-bit (f - x 1)
     (f 5)
'(1 0 1 0 1)

expression  (try 7 (' ((' (lambda (f) (f 5))) (' (lambda (x) (
            if (= 0 x) nil (cons (* 2 (display (read-bit))) (f
             (- x 1)))))))) (' (1 0 1 0 1)))
value       (success (2 0 2 0 2) (1 0 1 0 1))

try 7
'let (f x) if = 0 x nil
           cons * 2 display read-bit (f - x 1)
     (f 5)
'(1 0 1)

expression  (try 7 (' ((' (lambda (f) (f 5))) (' (lambda (x) (
            if (= 0 x) nil (cons (* 2 (display (read-bit))) (f
             (- x 1)))))))) (' (1 0 1)))
value       (failure out-of-data (1 0 1))

try no-time-limit
'let (f x) if = 0 x nil
           cons * 2 display read-bit (f - x 1)
     (f 5)
'(1 0 1)

expression  (try no-time-limit (' ((' (lambda (f) (f 5))) (' (
            lambda (x) (if (= 0 x) nil (cons (* 2 (display (re
            ad-bit))) (f (- x 1)))))))) (' (1 0 1)))
value       (failure out-of-data (1 0 1))

try 18
'let (f x) if = 0 x nil
           cons * 2 display read-bit (f - x 1)
     (f 16)
bits 'a

expression  (try 18 (' ((' (lambda (f) (f 16))) (' (lambda (x)
             (if (= 0 x) nil (cons (* 2 (display (read-bit)))
            (f (- x 1)))))))) (bits (' a)))
value       (success (0 2 2 0 0 0 0 2 0 0 0 0 2 0 2 0) (0 1 1
            0 0 0 0 1 0 0 0 0 1 0 1 0))

base10-to-2 128

expression  (base10-to-2 128)
value       (1 0 0 0 0 0 0 0)

base10-to-2 256

expression  (base10-to-2 256)
value       (1 0 0 0 0 0 0 0 0)

base10-to-2 257

expression  (base10-to-2 257)
value       (1 0 0 0 0 0 0 0 1)

base2-to-10 '(1 1 1 1)

expression  (base2-to-10 (' (1 1 1 1)))
value       15

base2-to-10 '(1 0 0 0 0)

expression  (base2-to-10 (' (1 0 0 0 0)))
value       16

base2-to-10 '(1 0 0 0 1)

expression  (base2-to-10 (' (1 0 0 0 1)))
value       17

try 20
'cons abcdef try 10
'let (f n) (f display + n 1) (f 0) [infinite loop]
nil nil

expression  (try 20 (' (cons abcdef (try 10 (' ((' (lambda (f)
             (f 0))) (' (lambda (n) (f (display (+ n 1)))))))
            nil))) nil)
value       (success (abcdef failure out-of-time (1 2 3 4 5 6
            7 8 9)) ())

try 10
'cons abcdef try 20
'let (f n) (f display + n 1) (f 0) [infinite loop]
nil nil

expression  (try 10 (' (cons abcdef (try 20 (' ((' (lambda (f)
             (f 0))) (' (lambda (n) (f (display (+ n 1)))))))
            nil))) nil)
value       (failure out-of-time ())

try no-time-limit
'cons abcdef try 20
'let (f n) (f display + n 1) (f 0) [infinite loop]
nil nil

expression  (try no-time-limit (' (cons abcdef (try 20 (' (('
            (lambda (f) (f 0))) (' (lambda (n) (f (display (+
            n 1))))))) nil))) nil)
value       (success (abcdef failure out-of-time (1 2 3 4 5 6
            7 8 9 10 11 12 13 14 15 16 17 18 19)) ())

try 10
'cons abcdef try no-time-limit
'let (f n) (f display + n 1) (f 0) [infinite loop]
nil nil

expression  (try 10 (' (cons abcdef (try no-time-limit (' (('
            (lambda (f) (f 0))) (' (lambda (n) (f (display (+
            n 1))))))) nil))) nil)
value       (failure out-of-time ())

read-bit

expression  (read-bit)
value       out-of-data

read-exp

expression  (read-exp)
value       out-of-data

bits '(abc def)

expression  (bits (' (abc def)))
value       (0 0 1 0 1 0 0 0 0 1 1 0 0 0 0 1 0 1 1 0 0 0 1 0 0
             1 1 0 0 0 1 1 0 0 1 0 0 0 0 0 0 1 1 0 0 1 0 0 0 1
             1 0 0 1 0 1 0 1 1 0 0 1 1 0 0 0 1 0 1 0 0 1 0 0 0
             0 1 0 1 0)

try no-time-limit 'read-exp bits '(abc def)

expression  (try no-time-limit (' (read-exp)) (bits (' (abc de
            f))))
value       (success (abc def) ())

bits 'abc

expression  (bits (' abc))
value       (0 1 1 0 0 0 0 1 0 1 1 0 0 0 1 0 0 1 1 0 0 0 1 1 0
             0 0 0 1 0 1 0)

'(abc (def ghi) j)

expression  (' (abc (def ghi) j))
value       (abc (def ghi) j)

try 0 'read-bit nil

expression  (try 0 (' (read-bit)) nil)
value       (failure out-of-data ())

try 0 'read-exp nil

expression  (try 0 (' (read-exp)) nil)
value       (failure out-of-data ())

try 0 'read-exp bits 'abc

expression  (try 0 (' (read-exp)) (bits (' abc)))
value       (success abc ())

try 0 'cons read-exp cons read-bit nil bits 'abc

expression  (try 0 (' (cons (read-exp) (cons (read-bit) nil)))
             (bits (' abc)))
value       (failure out-of-data ())

try 0 'cons read-exp cons read-bit nil append bits 'abc '(0)

expression  (try 0 (' (cons (read-exp) (cons (read-bit) nil)))
             (append (bits (' abc)) (' (0))))
value       (success (abc 0) ())

try 0 'cons read-exp cons read-bit nil append bits 'abc '(1)

expression  (try 0 (' (cons (read-exp) (cons (read-bit) nil)))
             (append (bits (' abc)) (' (1))))
value       (success (abc 1) ())

try 0 'read-exp bits '(a b)

expression  (try 0 (' (read-exp)) (bits (' (a b))))
value       (success (a b) ())

try 0 'cons read-exp cons read-bit nil bits '(a b)

expression  (try 0 (' (cons (read-exp) (cons (read-bit) nil)))
             (bits (' (a b))))
value       (failure out-of-data ())

try 0 'cons read-exp cons read-exp nil bits '(a b)

expression  (try 0 (' (cons (read-exp) (cons (read-exp) nil)))
             (bits (' (a b))))
value       (failure out-of-data ())

try 0 'read-exp bits '(abc(def ghi)j)

expression  (try 0 (' (read-exp)) (bits (' (abc (def ghi) j)))
            )
value       (success (abc (def ghi) j) ())

try 0 'read-exp '(1 1 1 1) [character is incomplete]

expression  (try 0 (' (read-exp)) (' (1 1 1 1)))
value       (failure out-of-data ())

try 0 'read-exp '(0 0 0 0 1 0 1 0) [nothing in record; only \n]

expression  (try 0 (' (read-exp)) (' (0 0 0 0 1 0 1 0)))
value       (success () ())

try 0 'cons read-exp cons read-exp nil append bits '(a b c) bits '(d e f)

expression  (try 0 (' (cons (read-exp) (cons (read-exp) nil)))
             (append (bits (' (a b c))) (bits (' (d e f)))))
value       (success ((a b c) (d e f)) ())

try 0 'read-exp '(1 1 1 1  1 1 1 1  0 0 0 0  1 0 1 0) [invalid character]

expression  (try 0 (' (read-exp)) (' (1 1 1 1 1 1 1 1 0 0 0 0
            1 0 1 0)))
value       (success () ())

= 0003 3

expression  (= 3 3)
value       true

000099

expression  99
value       99

x

expression  x
value       c

let x b x

expression  ((' (lambda (x) x)) b)
value       b

x

expression  x
value       c

let 99 45 99

expression  ((' (lambda (99) 99)) 45)
value       99

End of LISP Run

Elapsed time is 0 seconds.
\end{verbatim}
}\chap{godel.r}{\Size\begin{verbatim}
lisp.c

LISP Interpreter Run

[[[ Show that a formal system of lisp complexity H_lisp (FAS) = N
    cannot enable us to exhibit an elegant S-expression of
    size greater than N + 419.
    An elegant lisp expression is one with the property that no
    smaller S-expression has the same value.
    Setting: formal axiomatic system is never-ending lisp expression
    that displays elegant S-expressions.
]]]

[ Idea is to have a program P search for something X that can be proved
  to be more complex than P is, and therefore P can never find X.
  I.e., idea is to show that if this program halts we get a contradiction,
  and therefore the program doesn't halt. ]

define (size-it-and-run-it exp)
    cadr cons display size display exp
         cons eval exp
              nil

define      size-it-and-run-it
value       (lambda (exp) (car (cdr (cons (display (size (disp
            lay exp))) (cons (eval exp) nil)))))


(size-it-and-run-it'
+ 5 15
)

expression  (size-it-and-run-it (' (+ 5 15)))
display     (+ 5 15)
display     8
value       20


(size-it-and-run-it'

[ Examine list x for element that is more than n characters in size. ]
[ If not found returns false. ]
let (examine x n)
    if atom x  false
    if < n size car x  car x
    (examine cdr x n)

[ Here we are given the formal axiomatic system FAS. ]
let fas 'display ^ 10 439 [insert FAS here preceeded by ']

[ n = the number of characters in program including the FAS. ]
let n display + 419 size display fas  [ n = 419 + |FAS| ]

[ Loop running the formal axiomatic system ]
let (loop t)
  let v display try display t fas nil [Run the formal system for t time steps.]
  let s (examine caddr v n)   [Did it output an elegant s-exp larger than this
program?]
  if s eval s                 [If found elegant s-exp bigger than this program,
                               run it so that its output is our output
(contradiction!)]
  if = failure car v (loop + t 1) [If not, keep looping]
  failure                     [or halt if formal system halted.]

(loop 0)                      [Start loop running with t = 0.]

) [end size-it-and-run-it]

expression  (size-it-and-run-it (' ((' (lambda (examine) ((' (
            lambda (fas) ((' (lambda (n) ((' (lambda (loop) (l
            oop 0))) (' (lambda (t) ((' (lambda (v) ((' (lambd
            a (s) (if s (eval s) (if (= failure (car v)) (loop
             (+ t 1)) failure)))) (examine (car (cdr (cdr v)))
             n)))) (display (try (display t) fas nil)))))))) (
            display (+ 419 (size (display fas))))))) (' (displ
            ay (^ 10 439)))))) (' (lambda (x n) (if (atom x) f
            alse (if (< n (size (car x))) (car x) (examine (cd
            r x) n))))))))
display     ((' (lambda (examine) ((' (lambda (fas) ((' (lambd
            a (n) ((' (lambda (loop) (loop 0))) (' (lambda (t)
             ((' (lambda (v) ((' (lambda (s) (if s (eval s) (i
            f (= failure (car v)) (loop (+ t 1)) failure)))) (
            examine (car (cdr (cdr v))) n)))) (display (try (d
            isplay t) fas nil)))))))) (display (+ 419 (size (d
            isplay fas))))))) (' (display (^ 10 439)))))) (' (
            lambda (x n) (if (atom x) false (if (< n (size (ca
            r x))) (car x) (examine (cdr x) n))))))
display     439
display     (display (^ 10 439))
display     439
display     0
display     (success 10000000000000000000000000000000000000000
            00000000000000000000000000000000000000000000000000
            00000000000000000000000000000000000000000000000000
            00000000000000000000000000000000000000000000000000
            00000000000000000000000000000000000000000000000000
            00000000000000000000000000000000000000000000000000
            00000000000000000000000000000000000000000000000000
            00000000000000000000000000000000000000000000000000
            0000000000000000000000000000000000000000000000000
            (1000000000000000000000000000000000000000000000000
            00000000000000000000000000000000000000000000000000
            00000000000000000000000000000000000000000000000000
            00000000000000000000000000000000000000000000000000
            00000000000000000000000000000000000000000000000000
            00000000000000000000000000000000000000000000000000
            00000000000000000000000000000000000000000000000000
            00000000000000000000000000000000000000000000000000
            00000000000000000000000000000000000000000))
value       10000000000000000000000000000000000000000000000000
            00000000000000000000000000000000000000000000000000
            00000000000000000000000000000000000000000000000000
            00000000000000000000000000000000000000000000000000
            00000000000000000000000000000000000000000000000000
            00000000000000000000000000000000000000000000000000
            00000000000000000000000000000000000000000000000000
            00000000000000000000000000000000000000000000000000
            0000000000000000000000000000000000000000

End of LISP Run

Elapsed time is 0 seconds.
\end{verbatim}
}\chap{univtm.r}{\Size\begin{verbatim}
lisp.c

LISP Interpreter Run

[univtm.l]
[[[
 First steps with my new construction for
 a self-delimiting universal Turing machine.
 We show that
    H(x,y) <= H(x) + H(y) + c
 and determine c.
 Consider a bit string x of length |x|.
 We also show that
    H(x) <= 2|x| + c
 and that
    H(x) <= |x| + H(the binary string for |x|) + c
 and determine both these c's.
]]]

[first demo the new lisp primitive functions]
append '(1 2 3 4 5 6 7 8 9 0) '(a b c d e f g h i)

expression  (append (' (1 2 3 4 5 6 7 8 9 0)) (' (a b c d e f
            g h i)))
value       (1 2 3 4 5 6 7 8 9 0 a b c d e f g h i)

read-bit

expression  (read-bit)
value       out-of-data

try 0 'read-bit nil

expression  (try 0 (' (read-bit)) nil)
value       (failure out-of-data ())

try 0 'read-bit '(1)

expression  (try 0 (' (read-bit)) (' (1)))
value       (success 1 ())

try 0 'read-bit '(0)

expression  (try 0 (' (read-bit)) (' (0)))
value       (success 0 ())

try 0 'read-bit '(x)

expression  (try 0 (' (read-bit)) (' (x)))
value       (success 1 ())

try 0 'cons cons read-bit nil cons cons read-bit nil nil '(1 0)

expression  (try 0 (' (cons (cons (read-bit) nil) (cons (cons
            (read-bit) nil) nil))) (' (1 0)))
value       (success ((1) (0)) ())

try 0 'cons cons display read-bit nil cons cons display read-bit nil nil '(1 0)

expression  (try 0 (' (cons (cons (display (read-bit)) nil) (c
            ons (cons (display (read-bit)) nil) nil))) (' (1 0
            )))
value       (success ((1) (0)) (1 0))

try 0 'cons cons display read-bit nil cons cons display read-bit nil cons cons
display read-bit nil nil
      '(1 0)

expression  (try 0 (' (cons (cons (display (read-bit)) nil) (c
            ons (cons (display (read-bit)) nil) (cons (cons (d
            isplay (read-bit)) nil) nil)))) (' (1 0)))
value       (failure out-of-data (1 0))

try 0 'read-exp display bits a

expression  (try 0 (' (read-exp)) (display (bits a)))
display     (0 1 1 0 0 0 0 1 0 0 0 0 1 0 1 0)
value       (success a ())

try 0 'read-exp display bits b

expression  (try 0 (' (read-exp)) (display (bits b)))
display     (0 1 1 0 0 0 1 0 0 0 0 0 1 0 1 0)
value       (success b ())

try 0 'read-exp display bits c

expression  (try 0 (' (read-exp)) (display (bits c)))
display     (0 1 1 0 0 0 1 1 0 0 0 0 1 0 1 0)
value       (success c ())

try 0 'read-exp display bits d

expression  (try 0 (' (read-exp)) (display (bits d)))
display     (0 1 1 0 0 1 0 0 0 0 0 0 1 0 1 0)
value       (success d ())

try 0 'read-exp display bits e

expression  (try 0 (' (read-exp)) (display (bits e)))
display     (0 1 1 0 0 1 0 1 0 0 0 0 1 0 1 0)
value       (success e ())

try 0 'read-exp bits '(aa bb cc dd ee)

expression  (try 0 (' (read-exp)) (bits (' (aa bb cc dd ee))))
value       (success (aa bb cc dd ee) ())

try 0 'read-exp bits '(12 (3 4) 56)

expression  (try 0 (' (read-exp)) (bits (' (12 (3 4) 56))))
value       (success (12 (3 4) 56) ())

try 0 'cons read-exp cons read-exp nil
      append bits '(abc def) bits '(ghi jkl)

expression  (try 0 (' (cons (read-exp) (cons (read-exp) nil)))
             (append (bits (' (abc def))) (bits (' (ghi jkl)))
            ))
value       (success ((abc def) (ghi jkl)) ())

[
 Here is the self-delimiting universal Turing machine!
 (with slightly funny handling of out-of-tape condition)
]
define (U p) cadr try no-time-limit 'eval read-exp p

define      U
value       (lambda (p) (car (cdr (try no-time-limit (' (eval
            (read-exp))) p))))

[
 The length of this bit string is the
 constant c in H(x) <= 2|x| + 2 + c.
]
length bits '
let (loop) let x read-bit
           let y read-bit
           if = x y
              cons x (loop)
              nil
(loop)

expression  (length (bits (' ((' (lambda (loop) (loop))) (' (l
            ambda () ((' (lambda (x) ((' (lambda (y) (if (= x
            y) (cons x (loop)) nil))) (read-bit)))) (read-bit)
            )))))))
value       1104

(U
 append
   bits
   'let (loop) let x read-bit let y read-bit if = x y cons x (loop) nil
    (loop)

   '(0 0 1 1 0 0 1 1 0 1)
)

expression  (U (append (bits (' ((' (lambda (loop) (loop))) ('
             (lambda () ((' (lambda (x) ((' (lambda (y) (if (=
             x y) (cons x (loop)) nil))) (read-bit)))) (read-b
            it))))))) (' (0 0 1 1 0 0 1 1 0 1))))
value       (0 1 0 1)

(U
 append
   bits
   'let (loop) let x read-bit let y read-bit if = x y cons x (loop) nil
    (loop)

   '(0 0 1 1 0 0 1 1 0 0)
)

expression  (U (append (bits (' ((' (lambda (loop) (loop))) ('
             (lambda () ((' (lambda (x) ((' (lambda (y) (if (=
             x y) (cons x (loop)) nil))) (read-bit)))) (read-b
            it))))))) (' (0 0 1 1 0 0 1 1 0 0))))
value       out-of-data

[
 The length of this bit string is the
 constant c in H(x,y) <= H(x) + H(y) + c.
]
length bits '
cons eval read-exp
cons eval read-exp
     nil

expression  (length (bits (' (cons (eval (read-exp)) (cons (ev
            al (read-exp)) nil)))))
value       432

(U
 append
   bits 'cons eval read-exp cons eval read-exp nil
 append
   bits 'let (f) let x read-bit let y read-bit if = x y cons x (f) nil (f)
 append
   '(0 0 1 1 0 0 1 1 0 1)
 append
   bits 'let (f) let x read-bit let y read-bit if = x y cons x (f) nil (f)

   '(1 1 0 0 1 1 0 0 0 1)
)

expression  (U (append (bits (' (cons (eval (read-exp)) (cons
            (eval (read-exp)) nil)))) (append (bits (' ((' (la
            mbda (f) (f))) (' (lambda () ((' (lambda (x) ((' (
            lambda (y) (if (= x y) (cons x (f)) nil))) (read-b
            it)))) (read-bit))))))) (append (' (0 0 1 1 0 0 1
            1 0 1)) (append (bits (' ((' (lambda (f) (f))) ('
            (lambda () ((' (lambda (x) ((' (lambda (y) (if (=
            x y) (cons x (f)) nil))) (read-bit)))) (read-bit))
            ))))) (' (1 1 0 0 1 1 0 0 0 1)))))))
value       ((0 1 0 1) (1 0 1 0))

[
 The length of this bit string is the
 constant c in H(x) <= |x| + H(|x|) + c
]
length bits '
let (loop k)
   if = 0 k nil
   cons read-bit (loop - k 1)
(loop debug base2-to-10 eval debug read-exp)

expression  (length (bits (' ((' (lambda (loop) (loop (debug (
            base2-to-10 (eval (debug (read-exp)))))))) (' (lam
            bda (k) (if (= 0 k) nil (cons (read-bit) (loop (-
            k 1))))))))))
value       1152

(U
 append
   bits '
   let (loop k) if = 0 k nil cons read-bit (loop - k 1)
   (loop debug base2-to-10 eval debug read-exp)
 append
  bits ''(1 0 0 0) [Arbitrary program for U to compute number of bits]

   '(0 0 0 0  0 0 0 1) [that many bits of data]
)

expression  (U (append (bits (' ((' (lambda (loop) (loop (debu
            g (base2-to-10 (eval (debug (read-exp)))))))) (' (
            lambda (k) (if (= 0 k) nil (cons (read-bit) (loop
            (- k 1))))))))) (append (bits (' (' (1 0 0 0)))) (
            ' (0 0 0 0 0 0 0 1)))))
debug       (' (1 0 0 0))
debug       8
value       (0 0 0 0 0 0 0 1)

End of LISP Run

Elapsed time is 1 seconds.
\end{verbatim}
}\chap{godel2.r}{\Size

}\chap{omega.r}{\Size\begin{verbatim}
lisp.c

LISP Interpreter Run

[omega.l]

[[[[ Omega in the limit from below! ]]]]

[Generate all bit strings of length k]
define (all-bit-strings-of-size k)
    if = 0 k '(())
    (extend-by-one-bit (all-bit-strings-of-size - k 1))

define      all-bit-strings-of-size
value       (lambda (k) (if (= 0 k) (' (())) (extend-by-one-bi
            t (all-bit-strings-of-size (- k 1)))))

[Append 0 and 1 to each element of list]
define (extend-by-one-bit x)
    if atom x nil
    cons append car x '(0)
    cons append car x '(1)
    (extend-by-one-bit cdr x)

define      extend-by-one-bit
value       (lambda (x) (if (atom x) nil (cons (append (car x)
             (' (0))) (cons (append (car x) (' (1))) (extend-b
            y-one-bit (cdr x))))))

(extend-by-one-bit'((a)(b)))

expression  (extend-by-one-bit (' ((a) (b))))
value       ((a 0) (a 1) (b 0) (b 1))

(all-bit-strings-of-size 0)

expression  (all-bit-strings-of-size 0)
value       (())

(all-bit-strings-of-size 1)

expression  (all-bit-strings-of-size 1)
value       ((0) (1))

(all-bit-strings-of-size 2)

expression  (all-bit-strings-of-size 2)
value       ((0 0) (0 1) (1 0) (1 1))

(all-bit-strings-of-size 3)

expression  (all-bit-strings-of-size 3)
value       ((0 0 0) (0 0 1) (0 1 0) (0 1 1) (1 0 0) (1 0 1) (
            1 1 0) (1 1 1))

[Count programs in list p that halt within time t]
define (count-halt p t)
    if atom p 0
    +
    if = success display car try t 'eval debug read-exp car p
       1 0
    (count-halt cdr p t)

define      count-halt
value       (lambda (p t) (if (atom p) 0 (+ (if (= success (di
            splay (car (try t (' (eval (debug (read-exp)))) (c
            ar p))))) 1 0) (count-halt (cdr p) t))))

(count-halt cons bits '+ 10 15
            cons bits 'let(f)(f)(f)
                 nil
 99)

expression  (count-halt (cons (bits (' (+ 10 15))) (cons (bits
             (' ((' (lambda (f) (f))) (' (lambda () (f)))))) n
            il)) 99)
debug       (+ 10 15)
display     success
debug       ((' (lambda (f) (f))) (' (lambda () (f))))
display     failure
value       1

(count-halt cons append bits 'read-bit '(1)
            cons append bits 'read-exp '(1)
                 nil
 99)

expression  (count-halt (cons (append (bits (' (read-bit))) ('
             (1))) (cons (append (bits (' (read-exp))) (' (1))
            ) nil)) 99)
debug       (read-bit)
display     success
debug       (read-exp)
display     failure
value       1

[
 The k th lower bound on Omega
 is the number of k bit strings that halt on U within time k
 divided by 2 raised to the power k.
]
define (omega k) cons (count-halt (all-bit-strings-of-size k) k)
                 cons /
                 cons ^ 2 k
                      nil

define      omega
value       (lambda (k) (cons (count-halt (all-bit-strings-of-
            size k) k) (cons / (cons (^ 2 k) nil))))

(omega 0)

expression  (omega 0)
display     failure
value       (0 / 1)

(omega 1)

expression  (omega 1)
display     failure
display     failure
value       (0 / 2)

(omega 2)

expression  (omega 2)
display     failure
display     failure
display     failure
display     failure
value       (0 / 4)

(omega 3)

expression  (omega 3)
display     failure
display     failure
display     failure
display     failure
display     failure
display     failure
display     failure
display     failure
value       (0 / 8)

(omega 8)

expression  (omega 8)
display     failure
display     failure
display     failure
display     failure
display     failure
display     failure
display     failure
display     failure
display     failure
display     failure
debug       ()
display     success
display     failure
display     failure
display     failure
display     failure
display     failure
display     failure
display     failure
display     failure
display     failure
display     failure
display     failure
display     failure
display     failure
display     failure
display     failure
display     failure
display     failure
display     failure
display     failure
display     failure
display     failure
display     failure
display     failure
display     failure
display     failure
display     failure
display     failure
display     failure
display     failure
display     failure
display     failure
display     failure
display     failure
display     failure
display     failure
display     failure
display     failure
display     failure
display     failure
display     failure
display     failure
display     failure
display     failure
display     failure
display     failure
display     failure
display     failure
display     failure
display     failure
display     failure
display     failure
display     failure
display     failure
display     failure
display     failure
display     failure
display     failure
display     failure
display     failure
display     failure
display     failure
display     failure
display     failure
display     failure
display     failure
display     failure
display     failure
display     failure
display     failure
display     failure
display     failure
display     failure
display     failure
display     failure
display     failure
display     failure
display     failure
display     failure
display     failure
display     failure
display     failure
display     failure
display     failure
display     failure
display     failure
display     failure
display     failure
display     failure
display     failure
display     failure
display     failure
display     failure
display     failure
display     failure
display     failure
display     failure
display     failure
display     failure
display     failure
display     failure
display     failure
display     failure
display     failure
display     failure
display     failure
display     failure
display     failure
display     failure
display     failure
display     failure
display     failure
display     failure
display     failure
display     failure
display     failure
display     failure
display     failure
display     failure
display     failure
display     failure
display     failure
display     failure
display     failure
display     failure
display     failure
display     failure
display     failure
display     failure
display     failure
display     failure
display     failure
display     failure
display     failure
display     failure
display     failure
display     failure
display     failure
display     failure
display     failure
display     failure
display     failure
display     failure
display     failure
display     failure
display     failure
display     failure
display     failure
display     failure
display     failure
display     failure
display     failure
display     failure
display     failure
display     failure
display     failure
display     failure
display     failure
display     failure
display     failure
display     failure
display     failure
display     failure
display     failure
display     failure
display     failure
display     failure
display     failure
display     failure
display     failure
display     failure
display     failure
display     failure
display     failure
display     failure
display     failure
display     failure
display     failure
display     failure
display     failure
display     failure
display     failure
display     failure
display     failure
display     failure
display     failure
display     failure
display     failure
display     failure
display     failure
display     failure
display     failure
display     failure
display     failure
display     failure
display     failure
display     failure
display     failure
display     failure
display     failure
display     failure
display     failure
display     failure
display     failure
display     failure
display     failure
display     failure
display     failure
display     failure
display     failure
display     failure
display     failure
display     failure
display     failure
display     failure
display     failure
display     failure
display     failure
display     failure
display     failure
display     failure
display     failure
display     failure
display     failure
display     failure
display     failure
display     failure
display     failure
display     failure
display     failure
display     failure
display     failure
display     failure
display     failure
display     failure
display     failure
display     failure
display     failure
display     failure
display     failure
display     failure
display     failure
display     failure
display     failure
display     failure
display     failure
value       (1 / 256)

End of LISP Run

Elapsed time is 0 seconds.
\end{verbatim}
}\chap{omega2.r}{\Size\begin{verbatim}
lisp.c

LISP Interpreter Run

[omega2.l]

[[[[ Omega in the limit from below! ]]]]
[[[[ Version II ]]]]

[Count programs with prefix bit string p that halt within time t]
[among all possible extensions by e more bits]
define (count-halt prefix time bits-left-to-extend)
    if = bits-left-to-extend 0
    if = success display car try time 'eval debug read-exp display prefix
       1 0
    + (count-halt append prefix '(0) time - bits-left-to-extend 1)
      (count-halt append prefix '(1) time - bits-left-to-extend 1)

define      count-halt
value       (lambda (prefix time bits-left-to-extend) (if (= b
            its-left-to-extend 0) (if (= success (display (car
             (try time (' (eval (debug (read-exp)))) (display
            prefix))))) 1 0) (+ (count-halt (append prefix ('
            (0))) time (- bits-left-to-extend 1)) (count-halt
            (append prefix (' (1))) time (- bits-left-to-exten
            d 1)))))

(count-halt bits 'cons read-bit cons read-bit nil no-time-limit 0)

expression  (count-halt (bits (' (cons (read-bit) (cons (read-
            bit) nil)))) no-time-limit 0)
display     (0 0 1 0 1 0 0 0 0 1 1 0 0 0 1 1 0 1 1 0 1 1 1 1 0
             1 1 0 1 1 1 0 0 1 1 1 0 0 1 1 0 0 1 0 0 0 0 0 0 0
             1 0 1 0 0 0 0 1 1 1 0 0 1 0 0 1 1 0 0 1 0 1 0 1 1
             0 0 0 0 1 0 1 1 0 0 1 0 0 0 0 1 0 1 1 0 1 0 1 1 0
             0 0 1 0 0 1 1 0 1 0 0 1 0 1 1 1 0 1 0 0 0 0 1 0 1
             0 0 1 0 0 1 0 0 0 0 0 0 0 1 0 1 0 0 0 0 1 1 0 0 0
             1 1 0 1 1 0 1 1 1 1 0 1 1 0 1 1 1 0 0 1 1 1 0 0 1
             1 0 0 1 0 0 0 0 0 0 0 1 0 1 0 0 0 0 1 1 1 0 0 1 0
             0 1 1 0 0 1 0 1 0 1 1 0 0 0 0 1 0 1 1 0 0 1 0 0 0
             0 1 0 1 1 0 1 0 1 1 0 0 0 1 0 0 1 1 0 1 0 0 1 0 1
             1 1 0 1 0 0 0 0 1 0 1 0 0 1 0 0 1 0 0 0 0 0 0 1 1
             0 1 1 1 0 0 1 1 0 1 0 0 1 0 1 1 0 1 1 0 0 0 0 1 0
             1 0 0 1 0 0 1 0 1 0 0 1 0 0 0 0 1 0 1 0)
debug       (cons (read-bit) (cons (read-bit) nil))
display     failure
value       0

(count-halt bits 'cons read-bit cons read-bit nil no-time-limit 1)

expression  (count-halt (bits (' (cons (read-bit) (cons (read-
            bit) nil)))) no-time-limit 1)
display     (0 0 1 0 1 0 0 0 0 1 1 0 0 0 1 1 0 1 1 0 1 1 1 1 0
             1 1 0 1 1 1 0 0 1 1 1 0 0 1 1 0 0 1 0 0 0 0 0 0 0
             1 0 1 0 0 0 0 1 1 1 0 0 1 0 0 1 1 0 0 1 0 1 0 1 1
             0 0 0 0 1 0 1 1 0 0 1 0 0 0 0 1 0 1 1 0 1 0 1 1 0
             0 0 1 0 0 1 1 0 1 0 0 1 0 1 1 1 0 1 0 0 0 0 1 0 1
             0 0 1 0 0 1 0 0 0 0 0 0 0 1 0 1 0 0 0 0 1 1 0 0 0
             1 1 0 1 1 0 1 1 1 1 0 1 1 0 1 1 1 0 0 1 1 1 0 0 1
             1 0 0 1 0 0 0 0 0 0 0 1 0 1 0 0 0 0 1 1 1 0 0 1 0
             0 1 1 0 0 1 0 1 0 1 1 0 0 0 0 1 0 1 1 0 0 1 0 0 0
             0 1 0 1 1 0 1 0 1 1 0 0 0 1 0 0 1 1 0 1 0 0 1 0 1
             1 1 0 1 0 0 0 0 1 0 1 0 0 1 0 0 1 0 0 0 0 0 0 1 1
             0 1 1 1 0 0 1 1 0 1 0 0 1 0 1 1 0 1 1 0 0 0 0 1 0
             1 0 0 1 0 0 1 0 1 0 0 1 0 0 0 0 1 0 1 0 0)
debug       (cons (read-bit) (cons (read-bit) nil))
display     failure
display     (0 0 1 0 1 0 0 0 0 1 1 0 0 0 1 1 0 1 1 0 1 1 1 1 0
             1 1 0 1 1 1 0 0 1 1 1 0 0 1 1 0 0 1 0 0 0 0 0 0 0
             1 0 1 0 0 0 0 1 1 1 0 0 1 0 0 1 1 0 0 1 0 1 0 1 1
             0 0 0 0 1 0 1 1 0 0 1 0 0 0 0 1 0 1 1 0 1 0 1 1 0
             0 0 1 0 0 1 1 0 1 0 0 1 0 1 1 1 0 1 0 0 0 0 1 0 1
             0 0 1 0 0 1 0 0 0 0 0 0 0 1 0 1 0 0 0 0 1 1 0 0 0
             1 1 0 1 1 0 1 1 1 1 0 1 1 0 1 1 1 0 0 1 1 1 0 0 1
             1 0 0 1 0 0 0 0 0 0 0 1 0 1 0 0 0 0 1 1 1 0 0 1 0
             0 1 1 0 0 1 0 1 0 1 1 0 0 0 0 1 0 1 1 0 0 1 0 0 0
             0 1 0 1 1 0 1 0 1 1 0 0 0 1 0 0 1 1 0 1 0 0 1 0 1
             1 1 0 1 0 0 0 0 1 0 1 0 0 1 0 0 1 0 0 0 0 0 0 1 1
             0 1 1 1 0 0 1 1 0 1 0 0 1 0 1 1 0 1 1 0 0 0 0 1 0
             1 0 0 1 0 0 1 0 1 0 0 1 0 0 0 0 1 0 1 0 1)
debug       (cons (read-bit) (cons (read-bit) nil))
display     failure
value       0

(count-halt bits 'cons read-bit cons read-bit nil no-time-limit 2)

expression  (count-halt (bits (' (cons (read-bit) (cons (read-
            bit) nil)))) no-time-limit 2)
display     (0 0 1 0 1 0 0 0 0 1 1 0 0 0 1 1 0 1 1 0 1 1 1 1 0
             1 1 0 1 1 1 0 0 1 1 1 0 0 1 1 0 0 1 0 0 0 0 0 0 0
             1 0 1 0 0 0 0 1 1 1 0 0 1 0 0 1 1 0 0 1 0 1 0 1 1
             0 0 0 0 1 0 1 1 0 0 1 0 0 0 0 1 0 1 1 0 1 0 1 1 0
             0 0 1 0 0 1 1 0 1 0 0 1 0 1 1 1 0 1 0 0 0 0 1 0 1
             0 0 1 0 0 1 0 0 0 0 0 0 0 1 0 1 0 0 0 0 1 1 0 0 0
             1 1 0 1 1 0 1 1 1 1 0 1 1 0 1 1 1 0 0 1 1 1 0 0 1
             1 0 0 1 0 0 0 0 0 0 0 1 0 1 0 0 0 0 1 1 1 0 0 1 0
             0 1 1 0 0 1 0 1 0 1 1 0 0 0 0 1 0 1 1 0 0 1 0 0 0
             0 1 0 1 1 0 1 0 1 1 0 0 0 1 0 0 1 1 0 1 0 0 1 0 1
             1 1 0 1 0 0 0 0 1 0 1 0 0 1 0 0 1 0 0 0 0 0 0 1 1
             0 1 1 1 0 0 1 1 0 1 0 0 1 0 1 1 0 1 1 0 0 0 0 1 0
             1 0 0 1 0 0 1 0 1 0 0 1 0 0 0 0 1 0 1 0 0 0)
debug       (cons (read-bit) (cons (read-bit) nil))
display     success
display     (0 0 1 0 1 0 0 0 0 1 1 0 0 0 1 1 0 1 1 0 1 1 1 1 0
             1 1 0 1 1 1 0 0 1 1 1 0 0 1 1 0 0 1 0 0 0 0 0 0 0
             1 0 1 0 0 0 0 1 1 1 0 0 1 0 0 1 1 0 0 1 0 1 0 1 1
             0 0 0 0 1 0 1 1 0 0 1 0 0 0 0 1 0 1 1 0 1 0 1 1 0
             0 0 1 0 0 1 1 0 1 0 0 1 0 1 1 1 0 1 0 0 0 0 1 0 1
             0 0 1 0 0 1 0 0 0 0 0 0 0 1 0 1 0 0 0 0 1 1 0 0 0
             1 1 0 1 1 0 1 1 1 1 0 1 1 0 1 1 1 0 0 1 1 1 0 0 1
             1 0 0 1 0 0 0 0 0 0 0 1 0 1 0 0 0 0 1 1 1 0 0 1 0
             0 1 1 0 0 1 0 1 0 1 1 0 0 0 0 1 0 1 1 0 0 1 0 0 0
             0 1 0 1 1 0 1 0 1 1 0 0 0 1 0 0 1 1 0 1 0 0 1 0 1
             1 1 0 1 0 0 0 0 1 0 1 0 0 1 0 0 1 0 0 0 0 0 0 1 1
             0 1 1 1 0 0 1 1 0 1 0 0 1 0 1 1 0 1 1 0 0 0 0 1 0
             1 0 0 1 0 0 1 0 1 0 0 1 0 0 0 0 1 0 1 0 0 1)
debug       (cons (read-bit) (cons (read-bit) nil))
display     success
display     (0 0 1 0 1 0 0 0 0 1 1 0 0 0 1 1 0 1 1 0 1 1 1 1 0
             1 1 0 1 1 1 0 0 1 1 1 0 0 1 1 0 0 1 0 0 0 0 0 0 0
             1 0 1 0 0 0 0 1 1 1 0 0 1 0 0 1 1 0 0 1 0 1 0 1 1
             0 0 0 0 1 0 1 1 0 0 1 0 0 0 0 1 0 1 1 0 1 0 1 1 0
             0 0 1 0 0 1 1 0 1 0 0 1 0 1 1 1 0 1 0 0 0 0 1 0 1
             0 0 1 0 0 1 0 0 0 0 0 0 0 1 0 1 0 0 0 0 1 1 0 0 0
             1 1 0 1 1 0 1 1 1 1 0 1 1 0 1 1 1 0 0 1 1 1 0 0 1
             1 0 0 1 0 0 0 0 0 0 0 1 0 1 0 0 0 0 1 1 1 0 0 1 0
             0 1 1 0 0 1 0 1 0 1 1 0 0 0 0 1 0 1 1 0 0 1 0 0 0
             0 1 0 1 1 0 1 0 1 1 0 0 0 1 0 0 1 1 0 1 0 0 1 0 1
             1 1 0 1 0 0 0 0 1 0 1 0 0 1 0 0 1 0 0 0 0 0 0 1 1
             0 1 1 1 0 0 1 1 0 1 0 0 1 0 1 1 0 1 1 0 0 0 0 1 0
             1 0 0 1 0 0 1 0 1 0 0 1 0 0 0 0 1 0 1 0 1 0)
debug       (cons (read-bit) (cons (read-bit) nil))
display     success
display     (0 0 1 0 1 0 0 0 0 1 1 0 0 0 1 1 0 1 1 0 1 1 1 1 0
             1 1 0 1 1 1 0 0 1 1 1 0 0 1 1 0 0 1 0 0 0 0 0 0 0
             1 0 1 0 0 0 0 1 1 1 0 0 1 0 0 1 1 0 0 1 0 1 0 1 1
             0 0 0 0 1 0 1 1 0 0 1 0 0 0 0 1 0 1 1 0 1 0 1 1 0
             0 0 1 0 0 1 1 0 1 0 0 1 0 1 1 1 0 1 0 0 0 0 1 0 1
             0 0 1 0 0 1 0 0 0 0 0 0 0 1 0 1 0 0 0 0 1 1 0 0 0
             1 1 0 1 1 0 1 1 1 1 0 1 1 0 1 1 1 0 0 1 1 1 0 0 1
             1 0 0 1 0 0 0 0 0 0 0 1 0 1 0 0 0 0 1 1 1 0 0 1 0
             0 1 1 0 0 1 0 1 0 1 1 0 0 0 0 1 0 1 1 0 0 1 0 0 0
             0 1 0 1 1 0 1 0 1 1 0 0 0 1 0 0 1 1 0 1 0 0 1 0 1
             1 1 0 1 0 0 0 0 1 0 1 0 0 1 0 0 1 0 0 0 0 0 0 1 1
             0 1 1 1 0 0 1 1 0 1 0 0 1 0 1 1 0 1 1 0 0 0 0 1 0
             1 0 0 1 0 0 1 0 1 0 0 1 0 0 0 0 1 0 1 0 1 1)
debug       (cons (read-bit) (cons (read-bit) nil))
display     success
value       4

(count-halt bits 'cons read-bit cons read-bit nil no-time-limit 3)

expression  (count-halt (bits (' (cons (read-bit) (cons (read-
            bit) nil)))) no-time-limit 3)
display     (0 0 1 0 1 0 0 0 0 1 1 0 0 0 1 1 0 1 1 0 1 1 1 1 0
             1 1 0 1 1 1 0 0 1 1 1 0 0 1 1 0 0 1 0 0 0 0 0 0 0
             1 0 1 0 0 0 0 1 1 1 0 0 1 0 0 1 1 0 0 1 0 1 0 1 1
             0 0 0 0 1 0 1 1 0 0 1 0 0 0 0 1 0 1 1 0 1 0 1 1 0
             0 0 1 0 0 1 1 0 1 0 0 1 0 1 1 1 0 1 0 0 0 0 1 0 1
             0 0 1 0 0 1 0 0 0 0 0 0 0 1 0 1 0 0 0 0 1 1 0 0 0
             1 1 0 1 1 0 1 1 1 1 0 1 1 0 1 1 1 0 0 1 1 1 0 0 1
             1 0 0 1 0 0 0 0 0 0 0 1 0 1 0 0 0 0 1 1 1 0 0 1 0
             0 1 1 0 0 1 0 1 0 1 1 0 0 0 0 1 0 1 1 0 0 1 0 0 0
             0 1 0 1 1 0 1 0 1 1 0 0 0 1 0 0 1 1 0 1 0 0 1 0 1
             1 1 0 1 0 0 0 0 1 0 1 0 0 1 0 0 1 0 0 0 0 0 0 1 1
             0 1 1 1 0 0 1 1 0 1 0 0 1 0 1 1 0 1 1 0 0 0 0 1 0
             1 0 0 1 0 0 1 0 1 0 0 1 0 0 0 0 1 0 1 0 0 0 0)
debug       (cons (read-bit) (cons (read-bit) nil))
display     success
display     (0 0 1 0 1 0 0 0 0 1 1 0 0 0 1 1 0 1 1 0 1 1 1 1 0
             1 1 0 1 1 1 0 0 1 1 1 0 0 1 1 0 0 1 0 0 0 0 0 0 0
             1 0 1 0 0 0 0 1 1 1 0 0 1 0 0 1 1 0 0 1 0 1 0 1 1
             0 0 0 0 1 0 1 1 0 0 1 0 0 0 0 1 0 1 1 0 1 0 1 1 0
             0 0 1 0 0 1 1 0 1 0 0 1 0 1 1 1 0 1 0 0 0 0 1 0 1
             0 0 1 0 0 1 0 0 0 0 0 0 0 1 0 1 0 0 0 0 1 1 0 0 0
             1 1 0 1 1 0 1 1 1 1 0 1 1 0 1 1 1 0 0 1 1 1 0 0 1
             1 0 0 1 0 0 0 0 0 0 0 1 0 1 0 0 0 0 1 1 1 0 0 1 0
             0 1 1 0 0 1 0 1 0 1 1 0 0 0 0 1 0 1 1 0 0 1 0 0 0
             0 1 0 1 1 0 1 0 1 1 0 0 0 1 0 0 1 1 0 1 0 0 1 0 1
             1 1 0 1 0 0 0 0 1 0 1 0 0 1 0 0 1 0 0 0 0 0 0 1 1
             0 1 1 1 0 0 1 1 0 1 0 0 1 0 1 1 0 1 1 0 0 0 0 1 0
             1 0 0 1 0 0 1 0 1 0 0 1 0 0 0 0 1 0 1 0 0 0 1)
debug       (cons (read-bit) (cons (read-bit) nil))
display     success
display     (0 0 1 0 1 0 0 0 0 1 1 0 0 0 1 1 0 1 1 0 1 1 1 1 0
             1 1 0 1 1 1 0 0 1 1 1 0 0 1 1 0 0 1 0 0 0 0 0 0 0
             1 0 1 0 0 0 0 1 1 1 0 0 1 0 0 1 1 0 0 1 0 1 0 1 1
             0 0 0 0 1 0 1 1 0 0 1 0 0 0 0 1 0 1 1 0 1 0 1 1 0
             0 0 1 0 0 1 1 0 1 0 0 1 0 1 1 1 0 1 0 0 0 0 1 0 1
             0 0 1 0 0 1 0 0 0 0 0 0 0 1 0 1 0 0 0 0 1 1 0 0 0
             1 1 0 1 1 0 1 1 1 1 0 1 1 0 1 1 1 0 0 1 1 1 0 0 1
             1 0 0 1 0 0 0 0 0 0 0 1 0 1 0 0 0 0 1 1 1 0 0 1 0
             0 1 1 0 0 1 0 1 0 1 1 0 0 0 0 1 0 1 1 0 0 1 0 0 0
             0 1 0 1 1 0 1 0 1 1 0 0 0 1 0 0 1 1 0 1 0 0 1 0 1
             1 1 0 1 0 0 0 0 1 0 1 0 0 1 0 0 1 0 0 0 0 0 0 1 1
             0 1 1 1 0 0 1 1 0 1 0 0 1 0 1 1 0 1 1 0 0 0 0 1 0
             1 0 0 1 0 0 1 0 1 0 0 1 0 0 0 0 1 0 1 0 0 1 0)
debug       (cons (read-bit) (cons (read-bit) nil))
display     success
display     (0 0 1 0 1 0 0 0 0 1 1 0 0 0 1 1 0 1 1 0 1 1 1 1 0
             1 1 0 1 1 1 0 0 1 1 1 0 0 1 1 0 0 1 0 0 0 0 0 0 0
             1 0 1 0 0 0 0 1 1 1 0 0 1 0 0 1 1 0 0 1 0 1 0 1 1
             0 0 0 0 1 0 1 1 0 0 1 0 0 0 0 1 0 1 1 0 1 0 1 1 0
             0 0 1 0 0 1 1 0 1 0 0 1 0 1 1 1 0 1 0 0 0 0 1 0 1
             0 0 1 0 0 1 0 0 0 0 0 0 0 1 0 1 0 0 0 0 1 1 0 0 0
             1 1 0 1 1 0 1 1 1 1 0 1 1 0 1 1 1 0 0 1 1 1 0 0 1
             1 0 0 1 0 0 0 0 0 0 0 1 0 1 0 0 0 0 1 1 1 0 0 1 0
             0 1 1 0 0 1 0 1 0 1 1 0 0 0 0 1 0 1 1 0 0 1 0 0 0
             0 1 0 1 1 0 1 0 1 1 0 0 0 1 0 0 1 1 0 1 0 0 1 0 1
             1 1 0 1 0 0 0 0 1 0 1 0 0 1 0 0 1 0 0 0 0 0 0 1 1
             0 1 1 1 0 0 1 1 0 1 0 0 1 0 1 1 0 1 1 0 0 0 0 1 0
             1 0 0 1 0 0 1 0 1 0 0 1 0 0 0 0 1 0 1 0 0 1 1)
debug       (cons (read-bit) (cons (read-bit) nil))
display     success
display     (0 0 1 0 1 0 0 0 0 1 1 0 0 0 1 1 0 1 1 0 1 1 1 1 0
             1 1 0 1 1 1 0 0 1 1 1 0 0 1 1 0 0 1 0 0 0 0 0 0 0
             1 0 1 0 0 0 0 1 1 1 0 0 1 0 0 1 1 0 0 1 0 1 0 1 1
             0 0 0 0 1 0 1 1 0 0 1 0 0 0 0 1 0 1 1 0 1 0 1 1 0
             0 0 1 0 0 1 1 0 1 0 0 1 0 1 1 1 0 1 0 0 0 0 1 0 1
             0 0 1 0 0 1 0 0 0 0 0 0 0 1 0 1 0 0 0 0 1 1 0 0 0
             1 1 0 1 1 0 1 1 1 1 0 1 1 0 1 1 1 0 0 1 1 1 0 0 1
             1 0 0 1 0 0 0 0 0 0 0 1 0 1 0 0 0 0 1 1 1 0 0 1 0
             0 1 1 0 0 1 0 1 0 1 1 0 0 0 0 1 0 1 1 0 0 1 0 0 0
             0 1 0 1 1 0 1 0 1 1 0 0 0 1 0 0 1 1 0 1 0 0 1 0 1
             1 1 0 1 0 0 0 0 1 0 1 0 0 1 0 0 1 0 0 0 0 0 0 1 1
             0 1 1 1 0 0 1 1 0 1 0 0 1 0 1 1 0 1 1 0 0 0 0 1 0
             1 0 0 1 0 0 1 0 1 0 0 1 0 0 0 0 1 0 1 0 1 0 0)
debug       (cons (read-bit) (cons (read-bit) nil))
display     success
display     (0 0 1 0 1 0 0 0 0 1 1 0 0 0 1 1 0 1 1 0 1 1 1 1 0
             1 1 0 1 1 1 0 0 1 1 1 0 0 1 1 0 0 1 0 0 0 0 0 0 0
             1 0 1 0 0 0 0 1 1 1 0 0 1 0 0 1 1 0 0 1 0 1 0 1 1
             0 0 0 0 1 0 1 1 0 0 1 0 0 0 0 1 0 1 1 0 1 0 1 1 0
             0 0 1 0 0 1 1 0 1 0 0 1 0 1 1 1 0 1 0 0 0 0 1 0 1
             0 0 1 0 0 1 0 0 0 0 0 0 0 1 0 1 0 0 0 0 1 1 0 0 0
             1 1 0 1 1 0 1 1 1 1 0 1 1 0 1 1 1 0 0 1 1 1 0 0 1
             1 0 0 1 0 0 0 0 0 0 0 1 0 1 0 0 0 0 1 1 1 0 0 1 0
             0 1 1 0 0 1 0 1 0 1 1 0 0 0 0 1 0 1 1 0 0 1 0 0 0
             0 1 0 1 1 0 1 0 1 1 0 0 0 1 0 0 1 1 0 1 0 0 1 0 1
             1 1 0 1 0 0 0 0 1 0 1 0 0 1 0 0 1 0 0 0 0 0 0 1 1
             0 1 1 1 0 0 1 1 0 1 0 0 1 0 1 1 0 1 1 0 0 0 0 1 0
             1 0 0 1 0 0 1 0 1 0 0 1 0 0 0 0 1 0 1 0 1 0 1)
debug       (cons (read-bit) (cons (read-bit) nil))
display     success
display     (0 0 1 0 1 0 0 0 0 1 1 0 0 0 1 1 0 1 1 0 1 1 1 1 0
             1 1 0 1 1 1 0 0 1 1 1 0 0 1 1 0 0 1 0 0 0 0 0 0 0
             1 0 1 0 0 0 0 1 1 1 0 0 1 0 0 1 1 0 0 1 0 1 0 1 1
             0 0 0 0 1 0 1 1 0 0 1 0 0 0 0 1 0 1 1 0 1 0 1 1 0
             0 0 1 0 0 1 1 0 1 0 0 1 0 1 1 1 0 1 0 0 0 0 1 0 1
             0 0 1 0 0 1 0 0 0 0 0 0 0 1 0 1 0 0 0 0 1 1 0 0 0
             1 1 0 1 1 0 1 1 1 1 0 1 1 0 1 1 1 0 0 1 1 1 0 0 1
             1 0 0 1 0 0 0 0 0 0 0 1 0 1 0 0 0 0 1 1 1 0 0 1 0
             0 1 1 0 0 1 0 1 0 1 1 0 0 0 0 1 0 1 1 0 0 1 0 0 0
             0 1 0 1 1 0 1 0 1 1 0 0 0 1 0 0 1 1 0 1 0 0 1 0 1
             1 1 0 1 0 0 0 0 1 0 1 0 0 1 0 0 1 0 0 0 0 0 0 1 1
             0 1 1 1 0 0 1 1 0 1 0 0 1 0 1 1 0 1 1 0 0 0 0 1 0
             1 0 0 1 0 0 1 0 1 0 0 1 0 0 0 0 1 0 1 0 1 1 0)
debug       (cons (read-bit) (cons (read-bit) nil))
display     success
display     (0 0 1 0 1 0 0 0 0 1 1 0 0 0 1 1 0 1 1 0 1 1 1 1 0
             1 1 0 1 1 1 0 0 1 1 1 0 0 1 1 0 0 1 0 0 0 0 0 0 0
             1 0 1 0 0 0 0 1 1 1 0 0 1 0 0 1 1 0 0 1 0 1 0 1 1
             0 0 0 0 1 0 1 1 0 0 1 0 0 0 0 1 0 1 1 0 1 0 1 1 0
             0 0 1 0 0 1 1 0 1 0 0 1 0 1 1 1 0 1 0 0 0 0 1 0 1
             0 0 1 0 0 1 0 0 0 0 0 0 0 1 0 1 0 0 0 0 1 1 0 0 0
             1 1 0 1 1 0 1 1 1 1 0 1 1 0 1 1 1 0 0 1 1 1 0 0 1
             1 0 0 1 0 0 0 0 0 0 0 1 0 1 0 0 0 0 1 1 1 0 0 1 0
             0 1 1 0 0 1 0 1 0 1 1 0 0 0 0 1 0 1 1 0 0 1 0 0 0
             0 1 0 1 1 0 1 0 1 1 0 0 0 1 0 0 1 1 0 1 0 0 1 0 1
             1 1 0 1 0 0 0 0 1 0 1 0 0 1 0 0 1 0 0 0 0 0 0 1 1
             0 1 1 1 0 0 1 1 0 1 0 0 1 0 1 1 0 1 1 0 0 0 0 1 0
             1 0 0 1 0 0 1 0 1 0 0 1 0 0 0 0 1 0 1 0 1 1 1)
debug       (cons (read-bit) (cons (read-bit) nil))
display     success
value       8

[
 The k th lower bound on Omega
 is the number of k bit strings that halt on U within time k
 divided by 2 raised to the power k.
]
define (omega k) cons (count-halt nil k k)
                 cons /
                 cons ^ 2 k
                      nil

define      omega
value       (lambda (k) (cons (count-halt nil k k) (cons / (co
            ns (^ 2 k) nil))))

(omega 0)

expression  (omega 0)
display     ()
display     failure
value       (0 / 1)

(omega 1)

expression  (omega 1)
display     (0)
display     failure
display     (1)
display     failure
value       (0 / 2)

(omega 2)

expression  (omega 2)
display     (0 0)
display     failure
display     (0 1)
display     failure
display     (1 0)
display     failure
display     (1 1)
display     failure
value       (0 / 4)

(omega 3)

expression  (omega 3)
display     (0 0 0)
display     failure
display     (0 0 1)
display     failure
display     (0 1 0)
display     failure
display     (0 1 1)
display     failure
display     (1 0 0)
display     failure
display     (1 0 1)
display     failure
display     (1 1 0)
display     failure
display     (1 1 1)
display     failure
value       (0 / 8)

End of LISP Run

Elapsed time is 1 seconds.
\end{verbatim}
}\chap{omega3.r}{\Size

}

\part{The Software}

\chap{lisp.m}{\Size\begin{verbatim}
(***** LISP.M *****)

(* lisp interpreter *)

getbit[] :=
Block[ {x},
 trouble = False; (* Mma bug bypass *)
 If[ atom@ tape, (trouble = True; Throw@ "out-of-data")];
 x = car@ tape;
 tape = cdr@ tape;
 If[ x === 0, 0, 1 ]
]

getchar[] := FromCharacterCode[
  128*getbit[] + 64*getbit[] + 32*getbit[] + 16*getbit[] +
  8*getbit[] + 4*getbit[] + 2*getbit[] + getbit[]
]

getrecord[] :=
Block[ { c, line = "", str },
 inputbuffer2 = {};
 While[ "\n" =!= (c = getchar[]),
 line = line <> c ];
 If[ trouble, Throw@ "out-of-data"]; (* Mma bug bypass *)
(* keep only printable ASCII codes *)
 line = FromCharacterCode@
 Cases[ ToCharacterCode@ line, n_Integer /; 32 <= n < 127 ];
 str = StringToStream@ line;
 inputbuffer2 = ReadList[str, Word, TokenWords->{"(", ")"}];
 Close@ str;
(* convert strings of digits to integers *)
 inputbuffer2 = If[DigitQ@#, ToExpression@#, #]& /@ inputbuffer2 ;
]

getexp[rparenokay_:False] :=
Block[ { w, d, l = {} },
(* supply unlimited number of )'s if tokens run out *)
 If[ inputbuffer2 == {}, w = ")",
 w = First@ inputbuffer2;
 inputbuffer2 = Rest@ inputbuffer2 ];
 Switch[
 w,
 ")", Return@ If[rparenokay,")",{}],
 "(",
(While[ ")" =!= (d = getexp[True]),
 AppendTo[l,d]
 ];
 Return@ l
),
 _, w
 ]
]

atom[x_] :=
 MatchQ[ x, {}|_String|_Integer ]
car[x_] :=
 If[ atom@ x, x, First@ x ]
cdr[x_] :=
 If[ atom@ x, x, Rest@ x ]
cons[x_,y_] :=
 If[ MatchQ[y,_String|_Integer], x, Prepend[y,x] ]

eval[e2_,a_,d2_] :=

Block[ {e = e2, d = d2, f, args, x, y, z},
 If[ MatchQ[e,_Integer], Return@ e ];
 If[ atom@ e, Block[ {names,values,pos},
   {names,values} = a;
   pos = Position[names,e,{1}];
   Return@ If[ pos == {}, e, values[[pos[[1,1]]]] ]
 ]];
 f = eval[car@ e,a,d];
 e = cdr@ e;
 Switch[
 f,
 "'", Return@ car@ e,
 "if", Return@
 If[
 eval[car@ e,a,d] =!= "false",
 eval[car@cdr@ e,a,d],
 eval[car@cdr@cdr@ e,a,d]
 ]
 ];
 args = eval[#,a,d]& /@ e;
 x = car@ args;
 y = car@cdr@ args;
 z = car@cdr@cdr@ args;
 Switch[
 f,
 "read-bit", Return@ getbit[],
 "read-exp", Return@ (getrecord[]; getexp[]),
 "bits", Return@ Flatten[ ( Rest@ IntegerDigits[256 + #, 2] )& /@
                            ToCharacterCode[ output@x <> "\n" ] ],
 "car", Return@ car@ x,
 "cdr", Return@ cdr@ x,
 "cons", Return@ cons[x,y],
 "size", Return@ StringLength@ output@ x,
 "length", Return@ Length@ x,
 "+", Return@ (nmb@x + nmb@y),
 "-", Return@ If[ nmb@x < nmb@y, 0, nmb@x - nmb@y ],
 "*", Return@ (nmb@x * nmb@y),
 "^", Return@ (nmb@x ^ nmb@y),
 "<", Return@ If[nmb@x < nmb@y, "true", "false"],
 ">", Return@ If[nmb@x > nmb@y, "true", "false"],
 ">=", Return@ If[nmb@x >= nmb@y, "true", "false"],
 "<=", Return@ If[nmb@x <= nmb@y, "true", "false"],
 "base10-to-2", Return@ IntegerDigits[nmb@x, 2],
 "base2-to-10", Return@ Fold[ (2 #1 + If[#2===0,0,1])&, 0, x ],
 "append", Return@ Join[ If[atom@x,{},x], If[atom@y,{},y] ],
 "atom", Return@ If[ atom@ x, "true", "false" ],
 "=", Return@ If[ x === y, "true", "false" ],
 "display", Return@ (AppendTo[out,x];
               If[ display, print[ "display", output@ x ] ];
               x),
 "debug", Return@ (print[ "debug", output@ x ];
               x)
 ];
 If[ d == 0, Throw@ "out-of-time" ];
 d--;
 Switch[
 f,
 "eval", Return@ eval[x,,d],
 "try", Return@
 Block[ {out = {}, tape = z, display = False, xx},
 If[ x === "no-time-limit", xx = Infinity, xx = nmb@ x ];
 If[ xx < d,
 Catch@ {"success",eval[y,,xx],out} //
 If[ # === "out-of-time", {"failure",#,out}, # ] & ,
 Catch@ {"success",eval[y,,d],out} //
 If[ # === "out-of-time", Throw@ #, # ] &
 ] //
 If[ # === "out-of-data", {"failure",#,out}, # ] & ] (* end block *)
 ]; (* end switch *)
 f = cdr@ f;
 eval[ car@cdr@ f, bind[car@ f,args,a], d ]
]

nmb[n_Integer] := n
nmb[_] := 0

bind[vars_?atom,args_,a_] :=
 a

bind[vars_,args_,a_] :=
Block[ {names,values,pos},
 {names,values} = bind[ cdr@ vars, cdr@ args, a ];
 pos = Position[names, car@ vars, {1}];
 {Prepend[Delete[names,pos], car@ vars],
  Prepend[Delete[values,pos], car@ args]}
]

eval[e_] :=
(
 out = tape = {};
 display = True;
 print[ "expression", output@ e ];
 Catch[ eval[ e, {names,defs}, Infinity ] ]
)

eval[e_,,d_] := eval[e,{{"nil"},{{}}},d]

run[fn_] := run[fn, "lisp.m", ".r"]

word2[]:=
Block[ {w,line,str},
While[
 inputbuffer == {},
 line = Read[i,Record];
 If[ line == EndOfFile, Abort[] ];
 Print@ line;
 WriteString[o,line,"\n"];
 (* keep only printable ASCII codes *)
 line = FromCharacterCode@
 Cases[ ToCharacterCode@ line, n_Integer /; 32 <= n < 127 ];
 str = StringToStream@ line;
 inputbuffer = ReadList[str, Word, TokenWords->{"(", ")", "[", "]", "'",
"\""}];
 Close@ str
];
w = First@ inputbuffer;
inputbuffer = Rest@ inputbuffer;
If[DigitQ@w, ToExpression@w, w]  (* convert string of digits to integer *)
]

word[] :=
Block[ {w},
While[ True,
 w = word2[];
 If[ w =!= "[", Return@ w ];
 While[ word[] =!= "]" ]
]
]

get[sexp_:False,rparenokay_:False] :=

Block[ {w = word[], d, l ={}, name, def, body, varlist},
 Switch[
 w,
 ")", Return@ If[rparenokay,")",{}],
 "(",
 While[ ")" =!= (d = get[sexp,True]),
 AppendTo[l,d]
 ];
 Return@ l
 ];
 If[ sexp, Return@ w ];
 Switch[
 w,
 "\"", get[True],
 "cadr",
 {"car",{"cdr",get[]}},
 "caddr",
 {"car",{"cdr",{"cdr",get[]}}},
 "let",
 {name,def,body} = {get[],get[],get[]};
 If[
 !MatchQ[name,{}|_String|_Integer],
 varlist = Rest@ name;
 name = First@ name;
 def = {"'",{"lambda",varlist,def}}
 ];
 {{"'",{"lambda",{name},body}},def},
 "read-bit"|"read-exp",
                   {w},
 "car"|"cdr"|"atom"|"'"|"display"|"eval"|"bits"|
"debug"|"length"|"size"|"base2-to-10"|"base10-to-2",
                   {w,get[]},
 "cons"|"="|"lambda"|"append"|"define"|"+"|"-"|"*"|"^"|"<"|">"|"<="|">=",
                   {w,get[],get[]},
 "if"|"let"|"try", {w,get[],get[],get[]},
 _, w
 ]
]

(* output S-exp *)
output2[x_String] := x<>" "
output2[x_Integer] := ToString[x]<>" "
output2[{x___}] :=
Block[ {s},
 s = StringJoin["(", output2 /@ {x}];
 If[ StringTake[s,-1] == " ", s = StringDrop[s,-1] ];
 s <> ") "
]
output[x_] := StringDrop[ output2@x ,-1 ]

blanks = StringJoin@ Table[" ",{12}]

print[x_,y_] := (print2[x,StringTake[y,50]];
 print["",StringDrop[y,50]]) /; StringLength[y] > 50
print[x_,y_] := print2[x,y]
print2[x_,y_] := print3[StringTake[x<>blanks,12]<>y]
print3[x_] := (Print[x]; WriteString[o,x,"\n"])

let[n_,d_] :=
(
 print[ "define", output@ n ];
 print[ "value", output@ d ];
 PrependTo[names,n];
 PrependTo[defs,d];
)

run[fn_,whoami_,outputsuffix_] :=
(
 inputbuffer = {};
 names = {"nil"}; defs = {{}};
 t0 = SessionTime[];
 o = OpenWrite[fn<>outputsuffix];
 i = OpenRead[fn<>".l"];
 print3["Start of "<>whoami<>" run of "<>fn<>".l"];
 print3@ "";
 CheckAbort[
 While[True,
(print3@ "";
 Replace[#,{
 {"define",{func_,vars___},def_} :> let[func,{"lambda",{vars},def}],
 {"define",var_,def_} :> let[var,eval@ def],
 _ :> print[ "value", output@ eval@ # ]
 }]
)& @ get[];
 print3@ ""
 ],
 ];
 print3@ StringForm[
 "Elapsed time `` seconds",
 Round[SessionTime[]-t0]
 ];
 Close@ i;
 Close@ o
)

runall := run /@ {"examples","godel","univtm","godel2",
                 "omega","omega2","omega3","godel3"}

$RecursionLimit = $IterationLimit = Infinity
SetOptions[$Output,PageWidth->Infinity];
\end{verbatim}
}\chap{lisp.c}{\Size\begin{verbatim}
/* lisp.c: high-speed LISP interpreter */

/*
   The storage required by this interpreter is 8 * 4 = 32 bytes times
   the symbolic constant SIZE, which is 32 * 1,000,000 =
   32 megabytes.  To run this interpreter in small machines,
   reduce the #define SIZE 1000000 below.

   To compile, type
      cc -O -olisp lisp.c
   To run interactively, type
      lisp
   To run with output on screen, type
      lisp <test.l
   To run with output in file, type
      lisp <test.l >test.r

   Reference:  Kernighan & Ritchie,
   The C Programming Language, Second Edition,
   Prentice-Hall, 1988.
*/

#include <stdio.h>
#include <time.h>

#define SIZE 1000000 /* numbers of nodes of tree storage */
#define nil 0 /* end of list marker */

long car[SIZE], cdr[SIZE]; /* tree storage */
short atom[SIZE]; /* is it an atom? */
short numb[SIZE]; /* is it a number? */
/* The following is only used for atoms */
long vlst[SIZE]; /* bindings of each atom */
long pname[SIZE]; /* print name of each atom = list of characters in reverse */
/* The following is only used for atoms that are the names of primitive
functions */
short pf_numb[SIZE]; /* primitive function number (for interpreter switch) */
short pf_args[SIZE]; /* number of arguments + 1 (for input parser) */

long obj_lst; /* list of all atoms (& every other token read except numbers) */

/* locations of atoms in tree storage */
long wrd_nil, wrd_true, wrd_false, wrd_define, wrd_let, wrd_lambda, wrd_quote,
wrd_if;
long wrd_car, wrd_cdr, wrd_cadr, wrd_caddr, wrd_eval, wrd_try;
long wrd_no_time_limit, wrd_out_of_time, wrd_out_of_data, wrd_success,
wrd_failure;
long left_bracket, right_bracket, left_paren, right_paren, double_quote;
long wrd_zero, wrd_one;

long next = 0; /* next free node */
long col = 0; /* column in each 50 character chunk of output
                 (preceeded by 12 char prefix) */
time_t time1; /* clock at start of execution */
time_t time2; /* clock at end of execution */
long turing_machine_tapes; /* stack of binary data for try's */
long display_enabled; /* stack of flags whether to capture displays or not */
long captured_displays; /* stack of stubs to collect captured displays on */
long q; /* for converting s-expressions into lists of bits */
long buffer2; /* buffer for converting lists of bits into s-expressions */
              /* contains list of all the words in an input record */

void initialize_atoms(void); /* initialize atoms */
long mk_atom(long number, char *name, long args); /* make an atom */
long mk_numb(long value); /* make an number */
long mk_string(char *p); /* make list of characters */
long eq_wrd(long x, long y); /* are two lists of characters equal ? */
long lookup_word(long x); /* look up word in object list ? */
long cons(long x, long y); /* get free node & stuff x & y in it */
long out(char *x, long y); /* output expression */
void out_lst(long x); /* output list */
void out_atm(long x); /* output atom */
void out_chr(long x); /* output character */
long in_word2(void); /* read word */
long in_word(void); /* read word - skip comments */
long in(long mexp, long rparenokay); /* input m-exp */
long only_digits(long x); /* check if list of characters are exclusively digits
*/
long ev(long e); /* initialize and evaluate expression */
long eval(long e, long d); /* evaluate expression */
long evalst(long e, long d); /* evaluate list of expressions */
void clean_env(void); /* clean environment */
void restore_env(void); /* restore unclean environment */
/* bind values of arguments to formal parameters */
void bind(long vars, long args);
long append(long x, long y); /* append two lists */
long eq(long x, long y); /* equal predicate */
long length(long x); /* number of elements in list */
long compare(long x, long y); /* compare two decimal numbers */
long add1(long x); /* add 1 to decimal number */
long sub1(long x); /* subtract 1 from decimal number */
long nmb(long x); /* pick-up decimal number from atom & convert non-number to
zero */
long remove_leading_zeros(long x); /* from reversed list of digits of decimal
number */
long addition(long x, long y, long carry_in); /* add two decimal numbers */
long multiplication(long x, long y); /* multiply two decimal numbers */
long exponentiation(long base, long exponent); /* base raised to the power
exponent */
long subtraction(long x, long y, long borrow_in); /* x - y assumes x >= y */
long base2_to_10(long x); /* convert bit string to decimal number */
long halve(long x); /* used to convert decimal number to bit string */
long base10_to_2(long x); /* convert decimal number to bit string */
long size(long x); /* number of characters in print representation */
long read_bit(void); /* read one square of Turing machine tape */
void write_chr(long x); /* convert character into 8 bits */
void write_atm(long x); /* convert atom into 8 bits per character */
void write_lst(long x); /* convert s-exp into list of bits */
long read_record(void); /* read record from Turing machine tape */
long read_char(void); /* read one character from Turing machine tape */
long read_word(void); /* read word from Turing machine tape */
long read_expr(long rparenokay); /* read s-exp from Turing machine tape */

main() /* lisp main program */
{
time1 = time(NULL); /* start timer */
printf("lisp.c\n\nLISP Interpreter Run\n");
initialize_atoms();

while (1) {
      long e, f, name, def;
      printf("\n");
      /* read lisp meta-expression, ) not okay */
      e = in(1,0);
      printf("\n");
      f = car[e];
      name = car[cdr[e]];
      def = car[cdr[cdr[e]]];
      if (f == wrd_define) {
      /* definition */
         if (atom[name]) {
         /* variable definition, e.g., define x '(a b c) */
            def = out("expression",def);
            def = ev(def);
         } /* end of variable definition */
         else          {
         /* function definition, e.g., define (F x y) cons x cons y nil */
            long var_list = cdr[name];
            name = car[name];
            def = cons(wrd_lambda,cons(var_list,cons(def,nil)));
         } /* end of function definition */
         out("define",name);
         out("value",def);
         /* new binding replaces old */
         car[vlst[name]] = def;
         continue;
      } /* end of definition */
      /* write corresponding s-expression */
      e = out("expression",e);
      /* evaluate expression */
      e = out("value",ev(e));
   }
}

void initialize_atoms(void) /* initialize atoms */
{
 long i;
 if ( nil != mk_atom(0,"()",0) ) {
    printf("nil != 0\n");
    exit(0); /* terminate execution */
 }
 wrd_nil = mk_atom(0,"nil",0);
 car[vlst[wrd_nil]] = nil; /* so that value of nil is () */
 wrd_true = mk_atom(0,"true",0);
 wrd_false = mk_atom(0,"false",0);
 wrd_no_time_limit = mk_atom(0,"no-time-limit",0);
 wrd_out_of_time = mk_atom(0,"out-of-time",0);
 wrd_out_of_data = mk_atom(0,"out-of-data",0);
 wrd_success = mk_atom(0,"success",0);
 wrd_failure = mk_atom(0,"failure",0);
 wrd_define = mk_atom(0,"define",3);
 wrd_let = mk_atom(0,"let",4);
 wrd_lambda = mk_atom(0,"lambda",3);
 wrd_cadr = mk_atom(0,"cadr",2);
 wrd_caddr = mk_atom(0,"caddr",2);
 wrd_quote = mk_atom(0,"'",2);
 wrd_if = mk_atom(0,"if",4);
 wrd_car = mk_atom(1,"car",2);
 wrd_cdr = mk_atom(2,"cdr",2);
 i = mk_atom(3,"cons",3);
 i = mk_atom(4,"atom",2);
 i = mk_atom(5,"=",3);
 i = mk_atom(6,"display",2);
 i = mk_atom(7,"debug",2);
 i = mk_atom(8,"append",3);
 i = mk_atom(9,"length",2);
 i = mk_atom(10,"<",3);
 i = mk_atom(11,">",3);
 i = mk_atom(12,"<=",3);
 i = mk_atom(13,">=",3);
 i = mk_atom(14,"+",3);
 i = mk_atom(15,"*",3);
 i = mk_atom(16,"^",3);
 i = mk_atom(17,"-",3);
 i = mk_atom(18,"base2-to-10",2);
 i = mk_atom(19,"base10-to-2",2);
 i = mk_atom(20,"size",2);
 i = mk_atom(21,"read-bit",1);
 i = mk_atom(22,"bits",2);
 i = mk_atom(23,"read-exp",1);
 wrd_eval = mk_atom(0,"eval",2);
 wrd_try = mk_atom(0,"try",4);
 left_bracket = mk_atom(0,"[",0);
 right_bracket = mk_atom(0,"]",0);
 left_paren = mk_atom(0,"(",0);
 right_paren = mk_atom(0,")",0);
 double_quote = mk_atom(0,"\"",0);
 wrd_zero = mk_numb(nil);
 wrd_one = mk_numb(cons('1',nil));
}

long mk_atom(long number, char *name, long args) /* make an atom */
{
 long a;
 a = cons(nil,nil); /* get an empty node */
 car[a] = cdr[a] = a; /* so that car & cdr of atom = atom */
 atom[a] = 1;
 numb[a] = 0;
 pname[a] = mk_string(name);
 pf_numb[a] = number;
 pf_args[a] = args;
 /* initially each atom evaluates to self */
 vlst[a] = cons(a,nil);
 /* put on object list */
 obj_lst = cons(a,obj_lst);
 return a;
}

long mk_numb(long value) /* make an number */
{ /* digits are in reverse order, and 0 has empty list of digits */
 long a;
 a = cons(nil,nil); /* get an empty node */
 car[a] = cdr[a] = a; /* so that car & cdr of atom = atom */
 atom[a] = 1;
 numb[a] = 1;
 pname[a] = value; /* must make 00099 into 99 and 000 into empty list of digits
*/
 /* if necessary before calling this routine (to avoid removing leading zeros
unnecessarily) */
 pf_numb[a] = 0;
 pf_args[a] = 0;
 vlst[a] = 0;
 /* do not put on object list ! */
 return a;
}

long mk_string(char *p) /* make list of characters */
{             /* in reverse order */
 long v = nil;
 while (*p != '\0')
     v = cons(*p++,v);
 return v;
}

long cons(long x, long y) /* get free node & stuff x & y in it */
{
 long z;

 /* if y is not a list, then cons is x */
 if ( y != nil && atom[y] ) return x;

 if (next >= SIZE) {
  printf("Storage overflow!\n");
  exit(0);
 }

 z = next++;
 car[z] = x;
 cdr[z] = y;
 atom[z] = 0;
 numb[z] = 0;
 pname[z] = 0;
 pf_numb[z] = 0;
 pf_args[z] = 0;
 vlst[z] = 0;

 return z;
}

long out(char *x, long y) /* output expression */
{
   printf("%-12s",x);
   col = 0; /* so can insert \n and 12 blanks
               every 50 characters of output */
   out_lst(y);
   printf("\n");
   return y;
}

void out_lst(long x) /* output list */
{
   if (numb[x] && pname[x] == nil) {out_chr('0'); return;} /* null list of
digits means zero */
   if (atom[x]) {out_atm(pname[x]); return;}
   out_chr('(');
   while (!atom[x]) {
      out_lst(car[x]);
      x = cdr[x];
      if (!atom[x]) out_chr(' ');
   }
   out_chr(')');
}

void out_atm(long x) /* output atom */
{
   if (x == nil) return;
   out_atm(cdr[x]); /* output characters in reverse order */
   out_chr(car[x]);
}

void out_chr(long x) /* output character */
{
   if (col++ == 50) {printf("\n%-12s"," ");  col = 1;}
   putchar(x);
}

long eq_wrd(long x, long y) /* are two lists of characters equal ? */
{
   if (x == nil) return y == nil;
   if (y == nil) return 0;
   if (car[x] != car[y]) return 0;
   return eq_wrd(cdr[x],cdr[y]);
}

long lookup_word(long x) /* is word in object list ? */
{
   long i = obj_lst;
   while (!atom[i]) {
       /* if word is already in object list, don't make a new atom */
       if (eq_wrd(pname[car[i]],x)) return car[i];
       i = cdr[i];
   }
   /* if word isn't in object list, make new atom & add it to object list */
   i = mk_atom(0,"",0); /* adds word to object list */
   pname[i] = x;
   return i;
}

long in_word2(void) { /* read word */
   static long buffer = nil; /* buffer with all the words in a line of input */
   long character, word, line, end_of_line, end_of_buffer;
   while ( buffer == nil ) { /* read in a line */
      line = end_of_line = cons(nil,nil); /* stub */
      do { /* read characters until '\n' */
         character = getchar();
         if (character == EOF) {
            time2 = time(NULL);
            printf(
            "End of LISP Run\n\nElapsed time is %.0f seconds.\n",
            difftime(time2,time1)
         /* on some systems, above line should instead be: */
         /* time2 - time1 */
            );
            exit(0); /* terminate execution */
         } /* end of if (character == EOF) */
         putchar(character);
         /* add character to end of line */
         end_of_line = cdr[end_of_line] = cons(character,nil);
      }  /* end of read characters until '\n' */
      while (character != '\n');
      line = cdr[line]; /* remove stub at beginning of line */
      /* break line into words at  ( ) [ ] ' " characters */
      buffer = end_of_buffer = cons(nil,nil); /* stub */
      word = nil;
      while ( line != nil ) {
         character = car[line];
         line = cdr[line];
         /* look for characters that break words */
         if ( character == ' ' || character == '\n' ||
              character == '(' || character == ')' ||
              character == '[' || character == ']' ||
              character == '\'' || character == '\"' )
         { /* add nonempty word to end of buffer */
           if ( word != nil )
           end_of_buffer = cdr[end_of_buffer] = cons(word,nil);
           word = nil;
           /* add break character to end of buffer */
           if ( character != ' ' && character != '\n' )
           end_of_buffer = cdr[end_of_buffer] = cons(cons(character,nil),nil);
         }
         else
         { /* add character to word (in reverse order) */
           /* keep only nonblank printable ASCII codes */
           if (32 < character && character < 127)
           word = cons(character,word);
         }
      } /* end while ( line != nil ) */
      buffer = cdr[buffer]; /* remove stub at beginning of buffer */
   } /* end of do while ( buffer == nil ) */
   /* if buffer nonempty, return first word in buffer */
   word = car[buffer];
   buffer = cdr[buffer];
   /* first check if word consists only of digits */
   if (only_digits(word)) word = mk_numb(remove_leading_zeros(word));
   /* also makes 00099 into 99 and 0000 into null */
   else word = lookup_word(word); /* look up word in object list */
   /* also does mk_atom and adds it to object list if necessary */
   return word;
}

long only_digits(long x) /* check if list of characters are exclusively digits
*/
{
   while (x != nil) {
      long digit = car[x];
      if (digit < '0' || digit > '9') return 0;
      x = cdr[x];
   }
   return 1;
}

long in_word(void) /* read word - skip comments */
{
   long w;
   while (1) {
      w = in_word2();
      if (w != left_bracket) return w;
      while (in_word() != right_bracket) ; /* comments may be nested */
   }
}

long in(long mexp, long rparenokay) /* input m-exp */
{
   long w = in_word(), first, last, next, name, def, body, var_lst, i ;
   if (w == right_paren) if (rparenokay) return w; else return nil;
   if (w == left_paren) { /* explicit list */
      first = last = cons(nil,nil);
      while ((next = in(mexp,1)) != right_paren)
         last = cdr[last] = cons(next,nil);
      return cdr[first];
   } /* end if (w == left_paren) */
   if (!mexp) return w; /* atom */
   if (w == double_quote) return in(0,0); /* s-exp */
   if (w == wrd_cadr) /* expand cadr */
      return
        cons(wrd_car,
        cons(cons(wrd_cdr,
             cons(in(1,0),
                  nil)),
             nil));
   if (w == wrd_caddr) /* expand caddr */
      return
        cons(wrd_car,
        cons(cons(wrd_cdr,
             cons(cons(wrd_cdr,
                  cons(in(1,0),
                       nil)),
                  nil)),
             nil));
   if (w == wrd_let) { /* expand let name def body  */
      name = in(1,0);
      def  = in(1,0);
      body = in(1,0);
      if (!atom[name]) { /* let (name var_lst) def body */
         var_lst = cdr[name];
         name = car[name];
         def = cons(wrd_quote,
               cons(cons(wrd_lambda,
                    cons(var_lst,
                    cons(def,
                         nil))),
                    nil));
      } /* end if (!atom[name]) */
      return /* let name def body */
      cons(cons(wrd_quote,
           cons(cons(wrd_lambda,
                cons(cons(name,
                          nil),
                cons(body,
                     nil))),
                nil)),
      cons(def,
           nil));
   } /* end if (w == wrd_let) */
   i = pf_args[w];
   if (i == 0) return w; /* normal atom */
   /* atom is a primitive function with i-1 arguments */
   first = last = cons(w,nil);
   while (--i > 0)
      last = cdr[last] = cons(in(1,0),nil);
   return first;
}

long ev(long e) /* initialize and evaluate expression */
{
 long stub, v;
 turing_machine_tapes = cons(nil,nil);
 display_enabled = cons(1,nil);
 captured_displays = cons(nil,nil);
 v = eval(e,wrd_no_time_limit);
 return (v < 0 ? -v : v);
}

long eval(long e, long d) /* evaluate expression */
{
/*
 e is expression to be evaluated
 d is permitted depth - decimal integer, or wrd_no_time_limit
*/
 long f, v, args, x, y, z, vars, body, var;

 if (numb[e]) return e;
 /* find current binding of atomic expression */
 if (atom[e]) return car[vlst[e]];

 f = eval(car[e],d); /* evaluate function */
 e = cdr[e]; /* remove function from list of arguments */
 if (f < 0) return f; /* function = error value? */

 if (f == wrd_quote) return car[e]; /* quote */

 if (f == wrd_if) { /* if then else */
    v = eval(car[e],d);
    e = cdr[e];
    if (v < 0) return v; /* error? */
    if (v == wrd_false) e = cdr[e];
    return eval(car[e],d);
 }

 args = evalst(e,d); /* evaluate list of arguments */
 if (args < 0) return args; /* error? */

 x = car[args]; /* pick up first argument */
 y = car[cdr[args]]; /* pick up second argument */
 z = car[cdr[cdr[args]]]; /* pick up third argument */

 switch (pf_numb[f]) {
 case 1: return car[x];
 case 2: return cdr[x];
 case 3: return cons(x,y);
 case 4: return (atom[x] ? wrd_true : wrd_false);
 case 5: return (eq(x,y) ? wrd_true : wrd_false);
 case 6: if (car[display_enabled]) return out("display",x);
         else {long stub, old_end, new_end;
               stub = car[captured_displays];
               old_end = car[stub];
               new_end = cons(x,nil);
               cdr[old_end] = new_end;
               car[stub] = new_end;
               return x;}
 case 7: return out("debug",x);
 case 8: return append((atom[x]?nil:x),(atom[y]?nil:y));
 case 9: return mk_numb(length(x));
 case 10: return (compare(nmb(x),nmb(y)) == '<' ? wrd_true : wrd_false);
 case 11: return (compare(nmb(x),nmb(y)) == '>' ? wrd_true : wrd_false);
 case 12: return (compare(nmb(x),nmb(y)) != '>' ? wrd_true : wrd_false); /* <=
*/
 case 13: return (compare(nmb(x),nmb(y)) != '<' ? wrd_true : wrd_false); /* >=
*/
 case 14: return mk_numb(addition(nmb(x),nmb(y),0)); /* no carry in initially
*/
 case 15: return mk_numb(multiplication(nmb(x),nmb(y)));
 case 16: return mk_numb(exponentiation(nmb(x),nmb(y)));
 case 17: if (compare(nmb(x),nmb(y)) != '>') return mk_numb(nil); /* y too big
to subtract from x */
          else return
mk_numb(remove_leading_zeros(subtraction(nmb(x),nmb(y),0)));
                                                                  /* no borrow
in initially */
 case 18: return mk_numb(base2_to_10(x)); /* convert bit string to decimal
number */
 case 19: return base10_to_2(nmb(x)); /* convert decimal number to bit string
*/
 case 20: return mk_numb(size(x)); /* size of print representation of x */
 case 21: return read_bit(); /* read one square of Turing machine tape */
          /* convert s-exp to list of bits */
 case 22: {v = q = cons(nil,nil); write_lst(x); write_chr('\n'); return
cdr[v];}
          /* read lisp s-expression from Turing machine tape, 8 bits per char
*/
 case 23: {v = read_record(); if (v < 0) return v; return read_expr(0);}
 } /* end switch (pf_numb[f]) */

 if (d != wrd_no_time_limit) {
    if (d == nil) return - wrd_out_of_time; /* depth exceeded -> error! */
    d = sub1(d); /* decrement depth */
 }

 if (f == wrd_eval) {
    clean_env(); /* clean environment */
    v = eval(x,d);
    restore_env(); /* restore unclean environment */
    return v;
 }

 if (f == wrd_try) {
    long stub, old_try_has_smaller_time_limit = 0; /* assume normal case, that
x < d */
    if (x != wrd_no_time_limit) x = nmb(x); /* convert s-exp into number */
    if (x == wrd_no_time_limit || (d != wrd_no_time_limit && compare(x,d) !=
'<')) {
       old_try_has_smaller_time_limit = 1;
       x = d; /* continue to use older more constraining time limit */
    }
    turing_machine_tapes = cons(z,turing_machine_tapes);
    display_enabled = cons(0,display_enabled);
    stub = cons(0,nil); /* stub to grow list on */
    car[stub] = stub; /* car of stub gives end of list */
    captured_displays = cons(stub,captured_displays);
    clean_env();
    v = eval(y,x);
    restore_env();
    turing_machine_tapes = cdr[turing_machine_tapes];
    display_enabled = cdr[display_enabled];
    stub = cdr[car[captured_displays]]; /* remove stub */
    captured_displays = cdr[captured_displays];
    if (old_try_has_smaller_time_limit && v == - wrd_out_of_time) return v;
    if (v < 0) return cons(wrd_failure,cons(-v,cons(stub,nil)));
    return cons(wrd_success,cons(v,cons(stub,nil)));
 }

 f = cdr[f];
 vars = car[f];
 f = cdr[f];
 body = car[f];

 bind(vars,args);

 v = eval(body,d);

 /* unbind */
 while (!atom[vars]) {
    var = car[vars];
    if (atom[var])
       vlst[var] = cdr[vlst[var]];
    vars = cdr[vars];
 }

 return v;
}

void clean_env(void) /* clean environment */
{
 long o = obj_lst, var;
 while (o != nil) {
    var = car[o];
    vlst[var] = cons(var,vlst[var]); /* everything eval's to self */
    o = cdr[o];
 }
 car[vlst[wrd_nil]] = nil; /* except that value of nil is () */
}

void restore_env(void) /* restore unclean environment */
{
 long o = obj_lst, var;
 while (o != nil) {
    var = car[o];
    if (cdr[vlst[var]] != nil)  /* was token read in by read-exp within a try
*/
       vlst[var] = cdr[vlst[var]];
    o = cdr[o];
 }
}

/* bind values of arguments to formal parameters */
void bind(long vars, long args)
{
 long var;
 if (atom[vars]) return;
 bind(cdr[vars],cdr[args]);
 var = car[vars];
 if (atom[var])
    vlst[var] = cons(car[args],vlst[var]);
}

long evalst(long e, long d) /* evaluate list of expressions */
{
 long x, y;
 if (e == nil) return nil;
 x = eval(car[e],d);
 if (x < 0) return x; /* error? */
 y = evalst(cdr[e],d);
 if (y < 0) return y; /* error? */
 return cons(x,y);
}

long append(long x, long y) /* append two lists */
{
 if (x == nil) return y;
 return cons(car[x],append(cdr[x],y));
}

long eq(long x, long y) /* equal predicate */
{
 if (x == y) return 1;
 if (numb[x] && numb[y]) return eq_wrd(pname[x],pname[y]);
 if (numb[x] || numb[y]) return 0;
 if (atom[x] || atom[y]) return 0;
 if (eq(car[x],car[y])) return eq(cdr[x],cdr[y]);
 return 0;
}

long length(long x) /* number of elements in list */
{
 if (atom[x]) return nil; /* is zero */
 return add1(length(cdr[x]));
}

long compare(long x, long y) /* compare two decimal numbers */
{
 long already_decided, digit1, digit2;
 if (x == nil && y == nil) return '=';
 if (x == nil && y != nil) return '<';
 if (x != nil && y == nil) return '>';
 already_decided = compare(cdr[x],cdr[y]);
 if (already_decided != '=') return already_decided;
 digit1 = car[x];
 digit2 = car[y];
 if (digit1 == digit2) return '=';
 if (digit1 <  digit2) return '<';
 if (digit1 >  digit2) return '>';
}

long add1(long x) /* add 1 to decimal number */
{
 long digit;
 if (x == nil) return cons('1',nil);
 digit = car[x];
 if (digit != '9') return cons(digit+1,cdr[x]);
 return cons('0',add1(cdr[x]));
}

long sub1(long x) /* subtract 1 from decimal number */
{
 long digit;
 if (x == nil) return x; /* 0 - 1 = 0 */
 digit = car[x];
 if (digit == '1' && cdr[x] == nil) return nil; /* 1 - 1 = 0 */
 if (digit != '0') return cons(digit-1,cdr[x]);
 return cons('9',sub1(cdr[x]));
}

long nmb(long x) /* pick-up decimal number from atom & convert non-number to
zero */
{
    if (numb[x]) return pname[x];
    return nil;
}

long remove_leading_zeros(long x) /* from reversed list of digits of decimal
number */
{
 long rest, digit;
 if (x == nil) return nil;
 digit = car[x];
 rest = remove_leading_zeros(cdr[x]);
 if (rest == nil && digit == '0') return nil;
 return cons(digit,rest);
}

long addition(long x, long y, long carry_in)
{
   long sum, digit1, digit2, rest1, rest2;
   if (x == nil && !carry_in) return y;
   if (y == nil && !carry_in) return x;
   if (x != nil) {digit1 = car[x]; rest1 = cdr[x];}
            else {digit1 = '0'; rest1 = nil;}
   if (y != nil) {digit2 = car[y]; rest2 = cdr[y];}
            else {digit2 = '0'; rest2 = nil;}
   sum = digit1 + digit2 + carry_in - '0';
   if (sum <= '9') return cons(sum,addition(rest1,rest2,0));
   return cons(sum-10,addition(rest1,rest2,1));
}

long subtraction(long x, long y, long borrow_in) /* x - y assumes x >= y */
{
   long difference, digit1, digit2, rest1, rest2;
   if (y == nil && !borrow_in) return x;
   if (x != nil) {digit1 = car[x]; rest1 = cdr[x];}
            else {digit1 = '0'; rest1 = nil;}
   if (y != nil) {digit2 = car[y]; rest2 = cdr[y];}
            else {digit2 = '0'; rest2 = nil;}
   difference = digit1 - digit2 - borrow_in + '0';
   if (difference >= '0') return cons(difference,subtraction(rest1,rest2,0));
   return cons(difference+10,subtraction(rest1,rest2,1));
}

long multiplication(long x, long y) /* goes faster if x is small */
{
   long sum = nil;
   if (y == nil) return nil; /* otherwise produces result 0000 */
   while (x != nil) {
      long digit = car[x];
      while (digit-- > '0') sum = addition(sum,y,0);
      x = cdr[x];
      y = cons('0',y); /* these are where bad decimal numbers are generated if
y is zero */
   }
   return sum;
}

long exponentiation(long base, long exponent)
{
   long product = cons('1',nil);
   while (exponent != nil) {
      product = multiplication(base,product); /* multiply faster if smaller
comes first */
      exponent = sub1(exponent);
   }
   return product;
}

long base2_to_10(long x) /* convert bit string to decimal number */
{
   long result = nil;
   while (!atom[x]) {
      long next_bit = car[x];
      x = cdr[x];
      if (!numb[next_bit] || pname[next_bit] != nil) next_bit = 1; else
next_bit = 0;
      result = addition(result,result,next_bit);
   }
   return result;
}

long halve(long x) /* used to convert decimal number to bit string */
{
   long digit, next_digit, rest, halve_digit;
   if (x == nil) return x; /* half of 0 is 0 */
   digit = car[x] - '0';
   x = cdr[x];
   rest = halve(x);
   if (x == nil) next_digit = 0; else next_digit = car[x] - '0';
   next_digit = next_digit%2; /* remainder when divided by 2 */
   halve_digit = '0' + (digit/2) + (5*next_digit);
   if (halve_digit != '0' || rest != nil) return cons(halve_digit,rest);
   return nil;
}

long base10_to_2(long x) /* convert decimal number to bit string */
{
   long bits = nil;
   while (x != nil) {
      long digit = car[x] - '0';
      bits = cons((digit%2 ? wrd_one : wrd_zero),bits);
      x = halve(x);
   }
   return bits;
}

long size(long x) /* number of characters in print representation */
{
   long sum = nil;
   if (numb[x] && pname[x] == nil) return add1(nil); /* number zero */
   if (atom[x]) return length(pname[x]);
   while (!atom[x]) {
      sum = addition(sum,size(car[x]),0);
      x = cdr[x];
      if (!atom[x]) sum = add1(sum); /* blank separator */
   }
   return add1(add1(sum)); /* open & close paren */
}

/* read one square of Turing machine tape */
long read_bit(void)
{
 long x, tape = car[turing_machine_tapes];
 if (atom[tape]) return - wrd_out_of_data; /* tape finished ! */
 x = car[tape];
 car[turing_machine_tapes] = cdr[tape];
 if (!numb[x] || pname[x] != nil) return wrd_one;
 return wrd_zero;
}

void write_chr(long x) /* convert character to list of 8 bits */
{
 q = cdr[q] = cons(( x & 128 ? wrd_one : wrd_zero ), nil);
 q = cdr[q] = cons(( x &  64 ? wrd_one : wrd_zero ), nil);
 q = cdr[q] = cons(( x &  32 ? wrd_one : wrd_zero ), nil);
 q = cdr[q] = cons(( x &  16 ? wrd_one : wrd_zero ), nil);
 q = cdr[q] = cons(( x &   8 ? wrd_one : wrd_zero ), nil);
 q = cdr[q] = cons(( x &   4 ? wrd_one : wrd_zero ), nil);
 q = cdr[q] = cons(( x &   2 ? wrd_one : wrd_zero ), nil);
 q = cdr[q] = cons(( x &   1 ? wrd_one : wrd_zero ), nil);
}

void write_lst(long x) /* convert s-exp to list of bits */
{
   if (numb[x] && pname[x] == nil) {write_chr('0'); return;} /* null list of
digits means zero */
   if (atom[x]) {write_atm(pname[x]); return;}
   write_chr('(');
   while (!atom[x]) {
      write_lst(car[x]);
      x = cdr[x];
      if (!atom[x]) write_chr(' ');
   }
   write_chr(')');
}

void write_atm(long x) /* convert atom to 8 bits per character */
{
   if (x == nil) return;
   write_atm(cdr[x]); /* output characters in reverse order */
   write_chr(car[x]);
}

/* read one character from Turing machine tape */
long read_char(void)
{
 long c, b, i = 8;
 c = 0;
 while (i-- > 0) {
    b = read_bit();
    if (b < 0) return b; /* error? */
    if (pname[b] != nil) b = 1; else b = 0;
    c = c + c + b;
 }
 return c;
}

long read_record(void) /* read record from Turing machine tape */
{  /* fill buffer2 with all the words in an input record */
   long character, word, line, end_of_line, end_of_buffer;
   line = end_of_line = cons(nil,nil); /* stub */
   do { /* read characters until '\n' */
      character = read_char();
      if (character < 0) return character; /* error? */;
      /* add character to end of line */
      end_of_line = cdr[end_of_line] = cons(character,nil);
   }  /* end of read characters until '\n' */
   while (character != '\n');
   line = cdr[line]; /* remove stub at beginning of line */
   /* break line into words at ( ) characters */
   buffer2 = end_of_buffer = cons(nil,nil); /* stub */
   word = nil;
   while ( line != nil ) {
      character = car[line];
      line = cdr[line];
      /* look for characters that break words */
      if ( character == ' ' || character == '\n' ||
           character == '(' || character == ')' )
      { /* add nonempty word to end of buffer */
        if ( word != nil )
        end_of_buffer = cdr[end_of_buffer] = cons(word,nil);
        word = nil;
        /* add break character to end of buffer */
        if ( character != ' ' && character != '\n' )
        end_of_buffer = cdr[end_of_buffer] = cons(cons(character,nil),nil);
      }
      else
      { /* add character to word (in reverse order) */
        /* keep only nonblank printable ASCII codes */
        if (32 < character && character < 127)
        word = cons(character,word);
      }
   } /* end while ( line != nil ) */
   buffer2 = cdr[buffer2]; /* remove stub at beginning of buffer */
}

long read_word(void) { /* read word from Turing machine tape */
   /* buffer2 has all the words in the input record */
   long word;
   /* (if buffer empty, returns as many right parens as needed) */
   if (buffer2 == nil) return right_paren;
   /* if buffer nonempty, return first word in buffer */
   word = car[buffer2];
   buffer2 = cdr[buffer2];
   /* first check if word consists only of digits */
   if (only_digits(word)) word = mk_numb(remove_leading_zeros(word));
   /* also makes 00099 into 99 and 0000 into null */
   else word = lookup_word(word); /* look up word in object list */
   /* also does mk_atom and adds it to object list if necessary */
   return word;
}

long read_expr(long rparenokay) /* read s-exp from Turing machine tape */
{
   long w = read_word(), first, last, next;
   if (w < 0) return w; /* error? */
   if (w == right_paren) if (rparenokay) return w; else return nil;
   if (w == left_paren) { /* explicit list */
      first = last = cons(nil,nil);
      while ((next = read_expr(1)) != right_paren) {
         if (next < 0) return next; /* error? */
         last = cdr[last] = cons(next,nil);
      }
      return cdr[first];
   } /* end if (w == left_paren) */
   return w; /* normal atom */
}
\end{verbatim}
}\chap{examples.l}{\Size\begin{verbatim}
[ test new lisp ]
' (ab c d)
'(ab   cd  )
car '(aa bb cc)
cdr '(aa bb cc)
cadr '(aa bb cc)
caddr '(aa bb cc)
cons '(aa bb cc) '(dd ee ff)
car aa
cdr aa
cons aa bb
("cons aa)
("cons '(aa) '(bb) '(cc))
let x a x
x
atom ' aa
atom '(aa)
if true x y
if false x y
if xxx x y
let (f x) if atom display x x (f car x)
 (f '(((a)b)c))
f
let (cat x y) if atom x y cons car x (cat cdr x y)
    (cat '(a b c) '(d e f))
cat
define (cat x y) if atom x y cons car x (cat cdr x y)
cat
(cat '(a b c) '(d e f))
define x cadr '(a b c)
x
define x caddr '(a b c)
x
length display
bits ' a
length display
bits ' abc
nil
length display
bits nil
length display
bits ' (a)
size abc
size ' ( a b c )
length ' ( a b c )
+ abc 15
+ '(abc) 15
+ 10 15
- 10 15
- 15 10
* 10 15
^ 10 15
< 10 15
< 10 10
> 10 15
> 10 10
<= 10 15
<= 10 10
>= 10 15
>= 10 10
= 10 15
= 10 10
let (f x) if = 0 x 1 * display x (f - x 1)
    (f 5)
let (f x) if = 0 x 1 * x (f - x 1)
    (f 100)
try 0
'let (f x) if = 0 x 1 * display x (f - x 1)
     (f 5)
nil
try 1
'let (f x) if = 0 x 1 * display x (f - x 1)
     (f 5)
nil
try 2
'let (f x) if = 0 x 1 * display x (f - x 1)
     (f 5)
nil
try 3
'let (f x) if = 0 x 1 * display x (f - x 1)
     (f 5)
nil
try 4
'let (f x) if = 0 x 1 * display x (f - x 1)
     (f 5)
nil
try 5
'let (f x) if = 0 x 1 * display x (f - x 1)
     (f 5)
nil
try 6
'let (f x) if = 0 x 1 * display x (f - x 1)
     (f 5)
nil
try 7
'let (f x) if = 0 x 1 * display x (f - x 1)
     (f 5)
nil
try no-time-limit
'let (f x) if = 0 x 1 * display x (f - x 1)
     (f 5)
nil
eval display '+ 5 15
try 6
'let (f x) if = 0 x nil
           cons * 2 display read-bit (f - x 1)
     (f 5)
'(1 0 1 0 1)
try 7
'let (f x) if = 0 x nil
           cons * 2 display read-bit (f - x 1)
     (f 5)
'(1 0 1 0 1)
try 7
'let (f x) if = 0 x nil
           cons * 2 display read-bit (f - x 1)
     (f 5)
'(1 0 1)
try no-time-limit
'let (f x) if = 0 x nil
           cons * 2 display read-bit (f - x 1)
     (f 5)
'(1 0 1)
try 18
'let (f x) if = 0 x nil
           cons * 2 display read-bit (f - x 1)
     (f 16)
bits 'a
base10-to-2 128
base10-to-2 256
base10-to-2 257
base2-to-10 '(1 1 1 1)
base2-to-10 '(1 0 0 0 0)
base2-to-10 '(1 0 0 0 1)
try 20
'cons abcdef try 10
'let (f n) (f display + n 1) (f 0) [infinite loop]
nil nil
try 10
'cons abcdef try 20
'let (f n) (f display + n 1) (f 0) [infinite loop]
nil nil
try no-time-limit
'cons abcdef try 20
'let (f n) (f display + n 1) (f 0) [infinite loop]
nil nil
try 10
'cons abcdef try no-time-limit
'let (f n) (f display + n 1) (f 0) [infinite loop]
nil nil
read-bit
read-exp
bits '(abc def)
try no-time-limit 'read-exp bits '(abc def)
bits 'abc
'(abc (def ghi) j)
try 0 'read-bit nil
try 0 'read-exp nil
try 0 'read-exp bits 'abc
try 0 'cons read-exp cons read-bit nil bits 'abc
try 0 'cons read-exp cons read-bit nil append bits 'abc '(0)
try 0 'cons read-exp cons read-bit nil append bits 'abc '(1)
try 0 'read-exp bits '(a b)
try 0 'cons read-exp cons read-bit nil bits '(a b)
try 0 'cons read-exp cons read-exp nil bits '(a b)
try 0 'read-exp bits '(abc(def ghi)j)
try 0 'read-exp '(1 1 1 1) [character is incomplete]
try 0 'read-exp '(0 0 0 0 1 0 1 0) [nothing in record; only \n]
try 0 'cons read-exp cons read-exp nil append bits '(a b c) bits '(d e f)
try 0 'read-exp '(1 1 1 1  1 1 1 1  0 0 0 0  1 0 1 0) [invalid character]
= 0003 3
000099
x
let x b x
x
let 99 45 99
\end{verbatim}
}\chap{godel.l}{\Size\begin{verbatim}
[[[ Show that a formal system of lisp complexity H_lisp (FAS) = N
    cannot enable us to exhibit an elegant S-expression of
    size greater than N + 419.
    An elegant lisp expression is one with the property that no
    smaller S-expression has the same value.
    Setting: formal axiomatic system is never-ending lisp expression
    that displays elegant S-expressions.
]]]

[ Idea is to have a program P search for something X that can be proved
  to be more complex than P is, and therefore P can never find X.
  I.e., idea is to show that if this program halts we get a contradiction,
  and therefore the program doesn't halt. ]

define (size-it-and-run-it exp)
    cadr cons display size display exp
         cons eval exp
              nil

(size-it-and-run-it'
+ 5 15
)

(size-it-and-run-it'

[ Examine list x for element that is more than n characters in size. ]
[ If not found returns false. ]
let (examine x n)
    if atom x  false
    if < n size car x  car x
    (examine cdr x n)

[ Here we are given the formal axiomatic system FAS. ]
let fas 'display ^ 10 439 [insert FAS here preceeded by ']

[ n = the number of characters in program including the FAS. ]
let n display + 419 size display fas  [ n = 419 + |FAS| ]

[ Loop running the formal axiomatic system ]
let (loop t)
  let v display try display t fas nil [Run the formal system for t time steps.]
  let s (examine caddr v n)   [Did it output an elegant s-exp larger than this
program?]
  if s eval s                 [If found elegant s-exp bigger than this program,
                               run it so that its output is our output
(contradiction!)]
  if = failure car v (loop + t 1) [If not, keep looping]
  failure                     [or halt if formal system halted.]

(loop 0)                      [Start loop running with t = 0.]

) [end size-it-and-run-it]
\end{verbatim}
}\chap{univtm.l}{\Size\begin{verbatim}
[univtm.l]
[[[
 First steps with my new construction for
 a self-delimiting universal Turing machine.
 We show that
    H(x,y) <= H(x) + H(y) + c
 and determine c.
 Consider a bit string x of length |x|.
 We also show that
    H(x) <= 2|x| + c
 and that
    H(x) <= |x| + H(the binary string for |x|) + c
 and determine both these c's.
]]]

[first demo the new lisp primitive functions]
append '(1 2 3 4 5 6 7 8 9 0) '(a b c d e f g h i)
read-bit
try 0 'read-bit nil
try 0 'read-bit '(1)
try 0 'read-bit '(0)
try 0 'read-bit '(x)
try 0 'cons cons read-bit nil cons cons read-bit nil nil '(1 0)
try 0 'cons cons display read-bit nil cons cons display read-bit nil nil '(1 0)
try 0 'cons cons display read-bit nil cons cons display read-bit nil cons cons
display read-bit nil nil
      '(1 0)
try 0 'read-exp display bits a
try 0 'read-exp display bits b
try 0 'read-exp display bits c
try 0 'read-exp display bits d
try 0 'read-exp display bits e
try 0 'read-exp bits '(aa bb cc dd ee)
try 0 'read-exp bits '(12 (3 4) 56)
try 0 'cons read-exp cons read-exp nil
      append bits '(abc def) bits '(ghi jkl)
[
 Here is the self-delimiting universal Turing machine!
 (with slightly funny handling of out-of-tape condition)
]
define (U p) cadr try no-time-limit 'eval read-exp p
[
 The length of this bit string is the
 constant c in H(x) <= 2|x| + 2 + c.
]
length bits '
let (loop) let x read-bit
           let y read-bit
           if = x y
              cons x (loop)
              nil
(loop)
(U
 append
   bits
   'let (loop) let x read-bit let y read-bit if = x y cons x (loop) nil
    (loop)

   '(0 0 1 1 0 0 1 1 0 1)
)
(U
 append
   bits
   'let (loop) let x read-bit let y read-bit if = x y cons x (loop) nil
    (loop)

   '(0 0 1 1 0 0 1 1 0 0)
)
[
 The length of this bit string is the
 constant c in H(x,y) <= H(x) + H(y) + c.
]
length bits '
cons eval read-exp
cons eval read-exp
     nil
(U
 append
   bits 'cons eval read-exp cons eval read-exp nil
 append
   bits 'let (f) let x read-bit let y read-bit if = x y cons x (f) nil (f)
 append
   '(0 0 1 1 0 0 1 1 0 1)
 append
   bits 'let (f) let x read-bit let y read-bit if = x y cons x (f) nil (f)

   '(1 1 0 0 1 1 0 0 0 1)
)
[
 The length of this bit string is the
 constant c in H(x) <= |x| + H(|x|) + c
]
length bits '
let (loop k)
   if = 0 k nil
   cons read-bit (loop - k 1)
(loop debug base2-to-10 eval debug read-exp)
(U
 append
   bits '
   let (loop k) if = 0 k nil cons read-bit (loop - k 1)
   (loop debug base2-to-10 eval debug read-exp)
 append
  bits ''(1 0 0 0) [Arbitrary program for U to compute number of bits]

   '(0 0 0 0  0 0 0 1) [that many bits of data]
)
\end{verbatim}
}\chap{godel2.l}{\Size\begin{verbatim}
[godel2.l]
[[[
 Show that a formal system of complexity N
 can't prove that a specific object has
 complexity > N + 4696.
 Formal system is a never halting lisp expression
 that output pairs (lisp object, lower bound
 on its complexity).  E.g., (x 4) means
 that x has complexity H(x) greater than or equal to 4.
]]]

[ Examine pairs to see if 2nd element is greater than lower bound. ]
[ Returns false to indicate not found, or pair if found. ]
define (examine pairs lower-bound)
    if atom pairs false
    if < lower-bound cadr car pairs
       car pairs
       (examine cdr pairs lower-bound)
(examine '((x 2)(y 3)) 0)
(examine '((x 2)(y 3)) 1)
(examine '((x 2)(y 3)) 2)
(examine '((x 2)(y 3)) 3)
(examine '((x 2)(y 3)) 4)

[This is an identity function with the size-effect of
displaying the number of bits in a binary string.]
define (display-number-of-bits string)
    cadr cons display length string cons string nil

cadr try no-time-limit 'eval read-exp [This is the universal Turing machine U
followed by its program]

[display number of bits in entire program]
(display-number-of-bits

append [append prefix and data]

[display number of bits in the prefix]
(display-number-of-bits bits '

[ Examine pairs to see if 2nd element is greater than lower bound. ]
[ Returns false to indicate not found, or pair if found. ]
let (examine pairs lower-bound)
    if atom pairs false
    if < lower-bound cadr car pairs
       car pairs
       (examine cdr pairs lower-bound)

[Main Loop - t is time limit, fas is bits of formal axiomatic system read so
far]
let (loop t fas)
 let v debug try debug t 'eval read-exp debug fas [run formal axiomatic system
again]
 [look for theorem which is pair with 2nd element > # of bits read + size of
this prefix]
 let s (examine caddr v debug + length fas 4696)
 if s car s       [Found it!  Output first element of theorem and halt]
 if = car v success failure [Surprise, formal system halts, so we do too]
 if = cadr v out-of-data  (loop t append fas cons read-bit nil)
                          [Read another bit of formal axiomatic system]
 if = cadr v out-of-time  (loop + t 1 fas)
                          [Increase time limit]
 unexpected-condition [This should never happen.]

(loop 0 nil)    [Initially, 0 time limit and no bits of formal axiomatic system
read]

) [end of prefix, start of formal axiomatic system]

bits ' display'(x 4881)

) [end of entire program for universal Turing machine U]
\end{verbatim}
}\chap{omega.l}{\Size\begin{verbatim}
[omega.l]

[[[[ Omega in the limit from below! ]]]]

[Generate all bit strings of length k]
define (all-bit-strings-of-size k)
    if = 0 k '(())
    (extend-by-one-bit (all-bit-strings-of-size - k 1))
[Append 0 and 1 to each element of list]
define (extend-by-one-bit x)
    if atom x nil
    cons append car x '(0)
    cons append car x '(1)
    (extend-by-one-bit cdr x)
(extend-by-one-bit'((a)(b)))
(all-bit-strings-of-size 0)
(all-bit-strings-of-size 1)
(all-bit-strings-of-size 2)
(all-bit-strings-of-size 3)
[Count programs in list p that halt within time t]
define (count-halt p t)
    if atom p 0
    +
    if = success display car try t 'eval debug read-exp car p
       1 0
    (count-halt cdr p t)
(count-halt cons bits '+ 10 15
            cons bits 'let(f)(f)(f)
                 nil
 99)
(count-halt cons append bits 'read-bit '(1)
            cons append bits 'read-exp '(1)
                 nil
 99)
[
 The k th lower bound on Omega
 is the number of k bit strings that halt on U within time k
 divided by 2 raised to the power k.
]
define (omega k) cons (count-halt (all-bit-strings-of-size k) k)
                 cons /
                 cons ^ 2 k
                      nil
(omega 0)
(omega 1)
(omega 2)
(omega 3)
(omega 8)
\end{verbatim}
}\chap{omega2.l}{\Size\begin{verbatim}
[omega2.l]

[[[[ Omega in the limit from below! ]]]]
[[[[ Version II ]]]]

[Count programs with prefix bit string p that halt within time t]
[among all possible extensions by e more bits]
define (count-halt prefix time bits-left-to-extend)
    if = bits-left-to-extend 0
    if = success display car try time 'eval debug read-exp display prefix
       1 0
    + (count-halt append prefix '(0) time - bits-left-to-extend 1)
      (count-halt append prefix '(1) time - bits-left-to-extend 1)
(count-halt bits 'cons read-bit cons read-bit nil no-time-limit 0)
(count-halt bits 'cons read-bit cons read-bit nil no-time-limit 1)
(count-halt bits 'cons read-bit cons read-bit nil no-time-limit 2)
(count-halt bits 'cons read-bit cons read-bit nil no-time-limit 3)
[
 The k th lower bound on Omega
 is the number of k bit strings that halt on U within time k
 divided by 2 raised to the power k.
]
define (omega k) cons (count-halt nil k k)
                 cons /
                 cons ^ 2 k
                      nil
(omega 0)
(omega 1)
(omega 2)
(omega 3)
\end{verbatim}
}\chap{omega3.l}{\Size\begin{verbatim}
[omega3.l]

[[[
 Show that
    H(Omega_n) > n - 9488.
 Omega_n is the first n bits of Omega,
 where we choose
    Omega = xxx0111111...
 instead of
    Omega = xxx1000000...
 if necessary.
]]]

[This is an identity function with the size-effect of
displaying the length in bits of the binary prefix.]
define (display-length-of-prefix prefix)
    cadr cons display length prefix cons prefix nil

cadr try no-time-limit 'eval read-exp [Univeral Turing machine U]

display
[followed by its program:]
append [append prefix and data]

[code to display size of prefix in bits]
(display-length-of-prefix bits '

let (count-halt prefix time bits-left-to-extend)
    if = bits-left-to-extend 0
    if = success car try time 'eval read-exp prefix
       1 0
    + (count-halt append prefix '(0) time - bits-left-to-extend 1)
      (count-halt append prefix '(1) time - bits-left-to-extend 1)

let (omega k) cons (count-halt nil k k)
              cons /
              cons ^ 2 k
                   nil

[Read and execute from remainder of tape
 a program to compute an n-bit
 initial piece of Omega.]
let w debug eval debug read-exp

[Convert to rational number]
let n length w
let w debug cons base2-to-10 w
            cons /
            cons ^ 2 n
                 nil

let (loop k)                         [Main Loop]
  let x debug (omega debug k)        [Compute the kth lower bound on Omega]
  if debug (<=rat w x) (big nil k n) [Are the first n bits OK?  If not, bump
k.]
     (loop + k 1) [Form the union of all output of n-bit
                   programs within time k, output it,
                   and halt.
                   This is bigger than anything of complexity
                   less than or equal to n!]
[This total output will be bigger than each individual output,
 and therefore must come from a program with more than n bits.
]

[Compare two rational binary numbers, i.e., is x = (a / b) <= y = (c / d) ?]
let (<=rat x y)
    let a car debug x
    let b caddr x
    let c car debug y
    let d caddr y
    <= * a d * b c

[Union of all output of n-bit programs within time k.]
let (big prefix time bits-left-to-add)
 if = 0 bits-left-to-add
 try time 'eval read-exp prefix
 append (big append prefix '(0) time - bits-left-to-add 1)
        (big append prefix '(1) time - bits-left-to-add 1)

(loop 0)         [Start main loop running with k = 0.]

)  [end of prefix]

bits '           [Here is the data: an optimal program to compute n bits of
Omega]

'(0 0 0 0  0 0 0 1)       [n = 8! Are these really the first 8 bits of Omega?]
\end{verbatim}
}\chap{godel3.l}{\Size\begin{verbatim}
[godel3.l]
[[[
 Show that a formal system of complexity N
 can't determine more than N + 9488 + 6912
 = N + 16400 bits of Omega.
 Formal system is a never halting lisp expression
 that outputs lists of the form (1 0 X 0 X X X X 1 0).
 This stands for the fractional part of Omega,
 and means that these 0,1 bits of Omega are known.
 X stands for an unknown bit.
]]]

[Count number of bits in an omega that are determined.]
define (number-of-bits-determined w)
    if atom w 0
    + (number-of-bits-determined cdr w)
      if = X car w
         0
         1
[Test it.]
(number-of-bits-determined '(X X X))
(number-of-bits-determined '(1 X X))
(number-of-bits-determined '(1 X 0))
(number-of-bits-determined '(1 1 0))

[Merge bits of data into unknown bits of an omega.]
define (supply-missing-bits w)
    if atom w nil
    cons if = X car w
            read-bit
            car w
    (supply-missing-bits cdr w)
[Test it.]
cadr try no-time-limit '
let (supply-missing-bits w)
    if atom w nil
    cons if = X car w
            read-bit
            car w
    (supply-missing-bits cdr w)
(supply-missing-bits '(0 0 X 0 0 X 0 0 X))
'(1 1 1)
cadr try no-time-limit '
let (supply-missing-bits w)
    if atom w nil
    cons if = X car w
            read-bit
            car w
    (supply-missing-bits cdr w)
(supply-missing-bits '(1 1 X 1 1 X 1 1 1))
'(0 0)

[
 Examine omegas in list w to see if in any one of them
 the number of bits that are determined is greater than n.
 Returns false to indicate not found, or what it found.
]
define (examine w n)
    if atom w false
    if < n (number-of-bits-determined car w)
       car w
       (examine cdr w n)
[Test it.]
(examine '((1 1)(1 1 1)) 0)
(examine '((1 1)(1 1 1)) 1)
(examine '((1 1)(1 1 1)) 2)
(examine '((1 1)(1 1 1)) 3)
(examine '((1 1)(1 1 1)) 4)

[This is an identity function with the size-effect of
displaying the number of bits in a binary string.]
define (display-number-of-bits string)
    cadr cons display length string
         cons string
              nil

cadr try no-time-limit 'eval read-exp [This is the universal Turing machine U
followed by its program]

append [Append missing bits of Omega to rest of program.]

[Display number of bits in entire program excepting the missing bits of Omega]
(display-number-of-bits

append [Append prefix and formal axiomatic system]

[Display number of bits in the prefix]
(display-number-of-bits bits '

[Count number of bits in an omega that are determined.]
let (number-of-bits-determined w)
    if atom w 0
    + (number-of-bits-determined cdr w)
      if = X car w
         0
         1

[Merge bits of data into unknown bits of an omega.]
let (supply-missing-bits w)
    if atom w nil
    cons if = X car w
            read-bit
            car w
    (supply-missing-bits cdr w)

[
 Examine omegas in list w to see if in any one of them
 the number of bits that are determined is greater than n.
 Return false to indicate not found, or what it found.
]
let (examine w n)
    if atom w false
[
    if < n (number-of-bits-determined car w)
]
    if < 1 (number-of-bits-determined car w)  [<==== changed n to 1 here so
will succeed]
       car w
       (examine cdr w n)

[Main Loop - t is time limit, fas is bits of formal axiomatic system read so
far]
let (loop t fas)
 let v debug try debug t 'eval read-exp debug fas
                              [Run formal axiomatic system again]
 [Look for theorem which determines more than (c + # of bits read + size of
this prefix)
  bits of Omega.  Here c = 9488 is the constant in the inequality in H(Omega_n)
> n - c
  (see omega3.l and omega3.r).]
 let s (examine caddr v + 9488 debug + length fas 6912)
 if s (supply-missing-bits s) [Found it!  Merge in undetermined bits, output
result, and halt.]
 if = car v success  failure  [Surprise, formal system halts, so we do too]
 if = cadr v out-of-data  (loop t append fas cons read-bit nil)
                              [Read another bit of formal axiomatic system]
 if = cadr v out-of-time  (loop + t 1 fas)
                              [Increase time limit]
 unexpected-condition         [This should never happen.]

(loop 0 nil)    [Initially, 0 time limit and no bits of formal axiomatic system
read]

) [end of prefix, start of formal axiomatic system]

[Toy formal system with only one theorem.]
bits 'display '(1 X 0)

) [end of prefix and formal axiomatic system]

'(1) [Missing bit of omega that is needed.]
\end{verbatim}
}

\end{document}